\documentclass[aps,pra,twocolumn,superscriptaddress,showpacs,amsmath,amssymb,nofootinbib,longbibliography]{revtex4-1}

\usepackage{graphicx}
\usepackage{hyperref}
\usepackage{array}
\usepackage{color}
\usepackage{multirow,here}

\newcommand{\ohm}{$\Omega$}
\newcommand{\Vout}{V_{\mathrm{OUT}}}
\newcommand{\Vin}{V_{\mathrm{IN}}}

\newcommand{\figtaskanalysis}{
\begin{figure}
\includegraphics{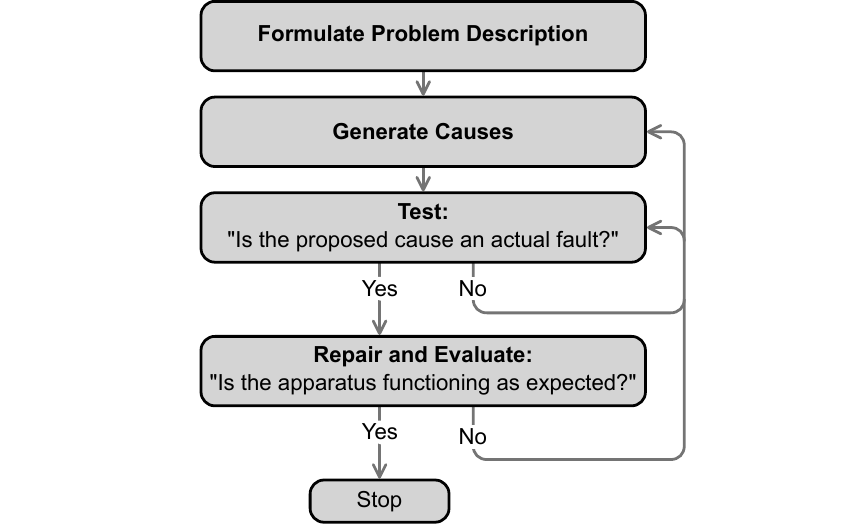}
\caption{\label{fig:taskanalysis} Cognitive troubleshooting tasks. These tasks describe the iterative process of repairing a malfunctioning apparatus. Performing each task requires up to six types of knowledge: domain, system, procedural, strategic, metacognitive, and experiential. This figure is based on the cognitive task analysis proposed by  Schaafstal \emph{et al.}~\cite{Schaafstal2000}.}
\end{figure}
}

\newcommand{\figmodelingframework}{
\begin{figure*}
\includegraphics{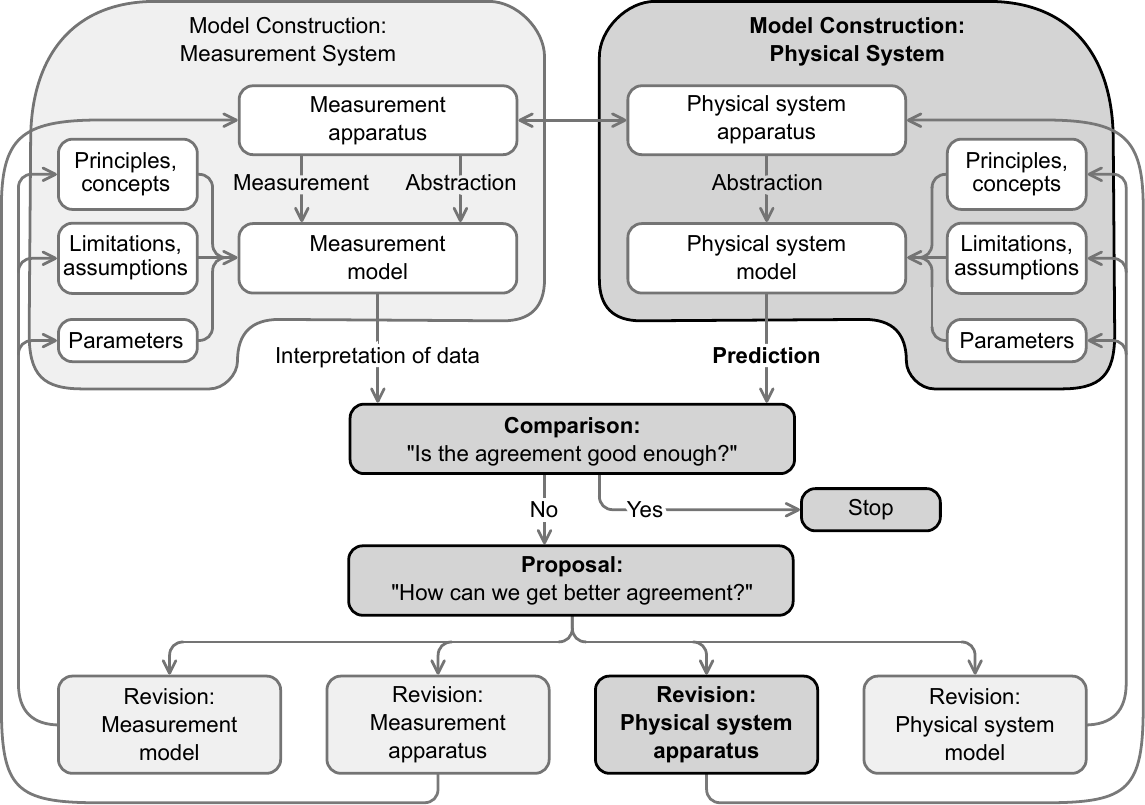}
\caption{\label{fig:modelingframework} Modeling Framework for Experimental Physics. This Framework describes the iterative process of constructing models of the measurement and physical systems, comparing measurements to predictions, proposing explanations for discrepancies, and revising models and/or apparatus. Darker shades of gray correspond to phases common in the troubleshooting process. Bold phrases indicate aspects of the Framework that informed our \emph{a priori} analysis scheme. This figure is adapted from the visualization presented by Zwickl \emph{et al.}~\cite{Zwickl2015}.}
\end{figure*}
}

\newcommand{\figdesign}{
\begin{figure*}
\includegraphics[width=\columnwidth]{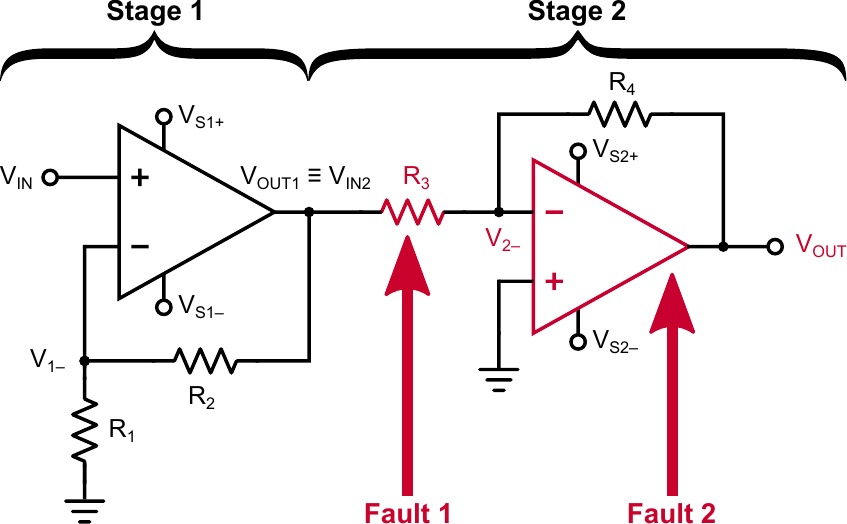}  \hfill \includegraphics[width=\columnwidth]{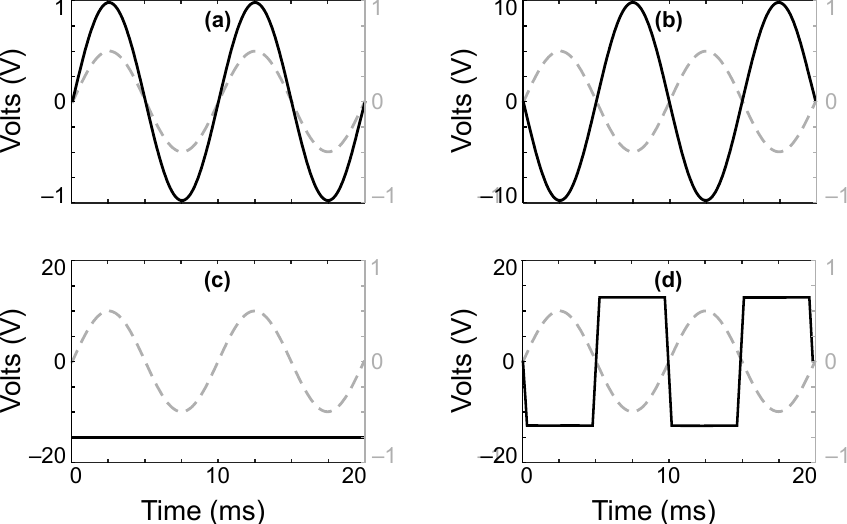}
\caption{\label{fig:design}(Left) Schematic diagram of inverting cascade amplifier, with design elements highlighted. A comparison of nominal voltages and resistances in both functioning and malfunctioning versions of the circuit is provided in Table~\ref{tab:faults}. (Right) Theoretical output signals of the inverting cascade amplifier with an input signal of amplitude 100~mV and frequency 100~Hz. In all plots, the vertical scales on the left and right respectively correspond to the output signal (solid black curve) and the input signal (dashed grey curve). Plots (a) and (b) show the outputs of Stages 1 and 2, respectively, in a functioning circuit. Plot (c) shows the output of Stage~2 due to the faulty op-amp, and Plot (d) shows the output of Stage~2 when both op-amps are functional but the input signal is sufficiently large that clipping occurs.}
\end{figure*}
}

\newcommand{\figphoto}{
\begin{figure}
\includegraphics[width=\columnwidth]{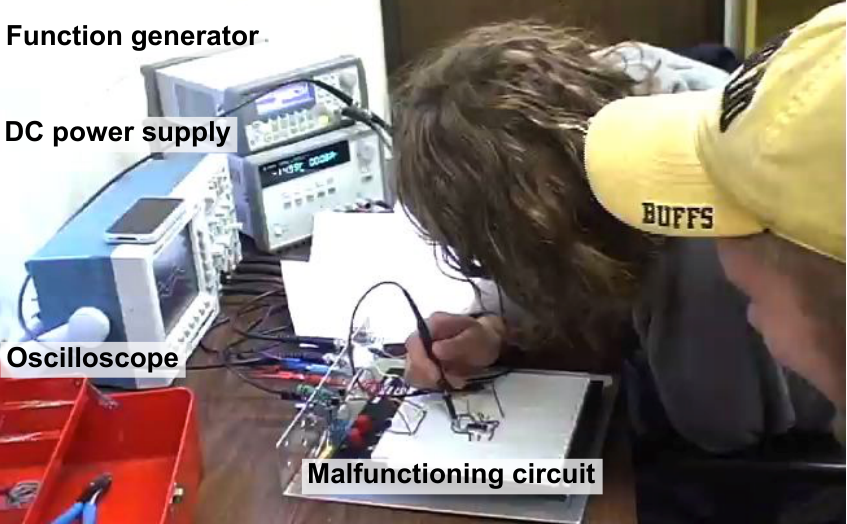}
\caption{\label{fig:photo}Photograph of the setup for the TAPPS interview.}
\end{figure}
}

\newcommand{\figschematic}{
\begin{figure}[t]
\includegraphics[width=\columnwidth]{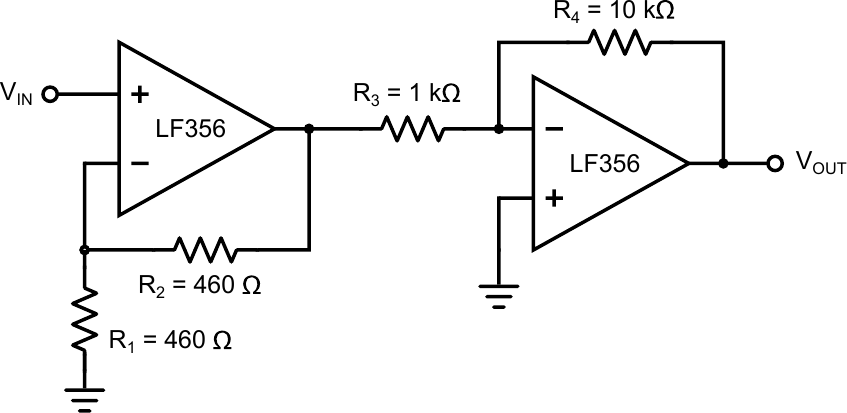}
\caption{\label{fig:schematic} Schematic diagram of inverting cascade amplifier, provided to study participants during the TAPPS activity. The diagram was accompanied by the following text: ``Figure 1 is a schematic of the inverting cascade amplifier we built. The [output] of this circuit is [given by Eq.~(\ref{eq:transfer})]. The main advantage of a cascade amplifier over a regular amplifier is that we can achieve high gain while maintaining a relatively large bandwidth. Disadvantages of this circuit compared to the regular amplifier include more components and increased power consumption."}
\end{figure}
}

\newcommand{\taskresults}{
\begin{figure*}
\includegraphics[width=\textwidth]{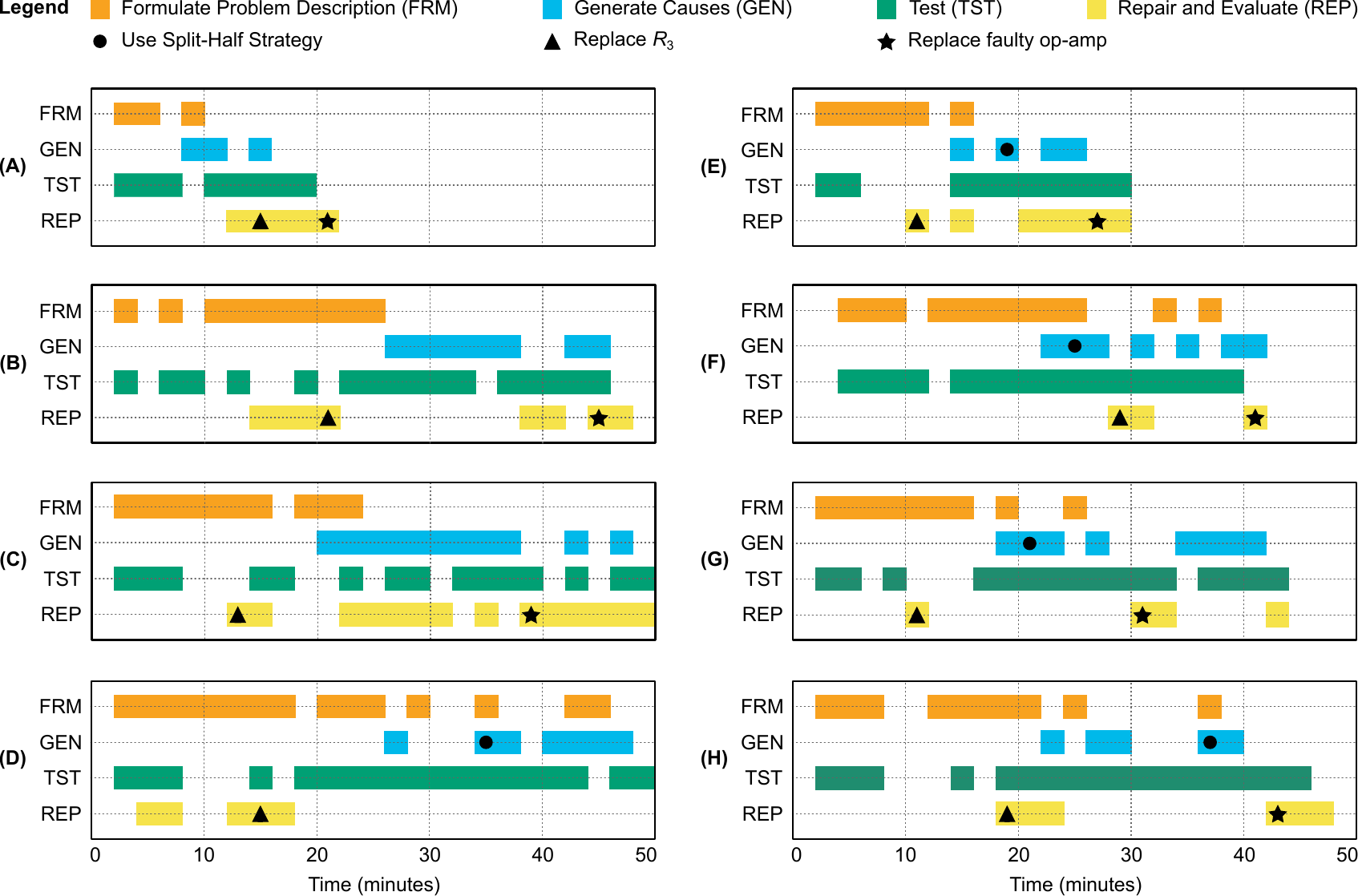}
\caption{\label{fig:tasks}Cognitive troubleshooting task coding results. Codes were applied to discrete two-minute time intervals. Colored bands indicate the times during which each pair engaged in the following tasks: Formulate Problem Description (FRM, orange); Generate Causes (GEN, blue); Test (TST, green); and Repair and Evaluate (REP, yellow). Filled black triangles and stars in the yellow REP bands indicate times at which students replaced the 100~\ohm\ resistor and faulty op-amp, respectively. Filled black circles in the blue GEN bands indicate times at which students employed the Split-Half Strategy. There are no codes in the first 2 minutes of the activity because the interviewer was giving verbal instructions during these times.}
\end{figure*}
}

\newcommand{\tabfaults}{
\begin{table}
\caption{\label{tab:faults} Nominal voltages and resistances in functioning (func.) and malfunctioning (mal.) versions of the circuit in Fig.~\ref{fig:design}, by stage. Characteristics of the malfunctioning circuit are listed only if they differ from the functioning case.}
\begin{ruledtabular}
\begin{tabular}{c>{$}l<{$}rr}
Stage & \mathrm{Quantity} & Value (func.) & Value (mal.) \\ \hline
\multirow{7}{*}{1} & \Vin & User defined \\
 & V_{\mathrm{S}1+} & $+15$ V dc \\
 & V_{\mathrm{S}1-} & $-15$ V dc \\
 & V_{1-} & $\Vin$ \\
 & V_{\mathrm{OUT}1} \equiv V_{\mathrm{IN}2} & $2\Vin$ \\
 & R_1 & 460 \ohm \\
 & R_2 & 460 \ohm \\ \hline
\multirow{7}{*}{2} & V_{\mathrm{IN}2} \equiv V_{\mathrm{OUT}1} & $2\Vin$ \\
 & V_{\mathrm{S}2+} & $+15$ V dc \\
 & V_{\mathrm{S}2-} & $-15$ V dc \\
 & V_{2-} & GND & $+15$ V dc \\
 & \Vout & $-10 V_{\mathrm{IN}2} = -20\Vin$ &  $-15$ V dc \\
 & R_3 & 1 k\ohm & 100 \ohm\\
 & R_4 & 10 k\ohm \\
\end{tabular}
\end{ruledtabular}
\end{table}
}

\newcommand{\taskcodes}{
\begin{table}
\caption{\label{tab:tasksubcodes} Codes and subcodes for cognitive troubleshooting tasks.}
\begin{ruledtabular}
\begin{tabular}{>{\bfseries}ll}
Code & Subcode \\ \hline
\multirow{3}{*}{\parbox{2cm}{\raggedright Formulate Problem Description}}
& Map circuit onto schematic and/or datasheet \\
& Discern functions of systems, components \\
& Perform formative measurements \\ \hline
\multirow{3}{*}{\parbox{2cm}{\raggedright Generate Causes}}
& Brainstorm potential causes or strategies\\
& Isolate subsystems as (mal)functioning \\
& General discussion about causes or strategies \\ \hline
\multirow{3}{*}{\parbox{2cm}{\raggedright Test}}
& Make a plan or prioritize measurements \\
& Formulate expectations about measurements \\
& Perform diagnostic measurements \\ \hline
\multirow{4}{*}{\parbox{2cm}{\raggedright Repair and Evaluate}}
& Propose a potential solution \\
& Replace component(s) \\
& Change circuit configuration \\
& Perform evaluative measurements
\end{tabular}
\end{ruledtabular}
\end{table}
}

\newcommand{\demotable}{
\begin{table}
\caption{\label{tab:demographics} Demographic breakdown of physics and engineering physics bachelor degree recipients at CU ($\sim47$ per year) and UM ($\sim11$ per year).}
\begin{ruledtabular}
\begin{tabular}{lrr}
Group & CU\footnote{Demographic data for 2005--14 were provided by the CU Office of Planning, Budget, and Analysis.} & UM\footnote{Demographic data for 2010--15 were provided by the UM Office of Institutional Research.} \\ \hline
Men & 84\% & 85\% \\
Women & 16\% & 15\% \\
White students & 77\% & 85\% \\
Asian students & 6\% & 2\% \\
URM students & 5\% & 4\% \\
Unspecified race/ethnicity & 10\% & 11\% \\
International students & 2\% & 2\%
\end{tabular}
\end{ruledtabular}
\end{table}
}

\newcommand{\modelingtable}{
\begin{table}
\caption{\label{tab:modelingcodes} Number and percent of groups who engaged in a particular aspect of the Modeling Framework in each of two troubleshooting episodes: isolating Stage 2 as the source of faults, and replacing the faulty op-amp. Percents were computed relative to the total number of groups who engaged in the corresponding troubleshooting episode.}
\begin{ruledtabular}
\begin{tabular}{l*{4}{r}}
& \multicolumn{2}{c}{Isolate Stage 2}
& \multicolumn{2}{c}{Replace Op-Amp} \\
& \multicolumn{2}{c}{$(N=5)$}
& \multicolumn{2}{c}{$(N=7)$} \\ \cline{2-3} \cline{4-5}
Code Category & Number & Percent & Number & Percent \\ \hline
Model Construction & 5 & 100\% & 4 & 57\% \\
Prediction & 5 & 100\% & 4 & 57\% \\
Comparison & 5 & 100\% & 7 & 100\% \\
Proposal & 2 & 40\% & 7 & 100\% \\
Revision & 0 & 0\% & 7 & 100\%
\end{tabular}
\end{ruledtabular}
\end{table}
}

\newcommand{\QUOTE}[4]{
\begin{quote} ``#1" \\ \rule{0pt}{0pt} \hfill (#2; #3--#4) \end{quote}
}

\begin{document}

\title{Investigating the role of model-based reasoning while troubleshooting an electric circuit}

\author{Dimitri R. Dounas-Frazer}
\email{dimitri.dounasfrazer@colorado.edu}
\affiliation{Department of Physics, University of Colorado Boulder, Boulder, CO 80309, USA}

\author{Kevin L. Van De Bogart}
\affiliation{Department of Physics and Astronomy, University of Maine, Orono, ME 04469, USA}

\author{MacKenzie R. Stetzer}
\affiliation{Department of Physics and Astronomy, University of Maine, Orono, ME 04469, USA}
\affiliation{Maine Center for Research in STEM Education, University of Maine, Orono, ME 04469, USA}

\author{H. J. Lewandowski}
\affiliation{Department of Physics, University of Colorado Boulder, Boulder, CO 80309, USA}
\affiliation{JILA, National Institute of Standards and Technology and University of Colorado Boulder, Boulder, CO 80309, USA}


\date{\today}

\begin{abstract}
We explore the overlap of two nationally-recognized learning outcomes for physics lab courses, namely, the ability to model experimental systems and the ability to troubleshoot a malfunctioning apparatus. Modeling and troubleshooting are both nonlinear, recursive processes that involve using models to inform revisions to an apparatus. To probe the overlap of modeling and troubleshooting, we collected audiovisual data from think-aloud activities in which eight pairs of students from two institutions attempted to diagnose and repair a malfunctioning electrical circuit. We characterize the cognitive tasks and model-based reasoning that students employed during this activity. In doing so, we demonstrate that troubleshooting engages students in the core scientific practice of modeling.
\end{abstract}

\maketitle

\pagebreak


\section{Introduction}

Recently, there have been national calls to study~\cite{DBER2012} and improve~\cite{PCAST2012} lab instruction in the sciences. Along these lines, the American Association of Physics Teachers (AAPT) released guidelines for learning outcomes in undergraduate physics lab courses~\cite{AAPT2015}. The AAPT guidelines focus on skill-based learning outcomes that align with the cognitive tasks involved in, for example, tabletop experimental physics research~\cite{Wieman2015}. In the present work, we investigate the overlap of two major learning goals of instructional physics lab environments: (1) the ability to model experimental systems, and (2) the ability to troubleshoot a malfunctioning apparatus. While these two abilities are sometimes presented as distinct, we aim to show that they are in fact overlapping both in theory and in practice. We show that, for some students, model-based reasoning plays a key role in the troubleshooting process.

Modeling is the nonlinear, recursive process of constructing, testing, and refining models~\cite{Zwickl2015}. Modeling has been identified as an important physics practice at both secondary~\cite{NGSS2013} and post-secondary levels~\cite{AAPT2015}. While traditional introductory physics lab courses have been criticized as rote and inauthentic~\cite{Wieman2015,NRC2013}, there nevertheless exist innovative approaches that engage students in the iterative process of modeling at the introductory level, such as ISLE~\cite{Etkina2007} and Modeling Instruction~\cite{Brewe2008}. At the upper-division level, the Modeling Framework for Experimental Physics (hereafter, ``the Modeling Framework") has been developed to characterize students' model-based reasoning~\cite{Zwickl2015} and to inform development of instructional lab environments that engage students in the practice of modeling~\cite{Zwickl2012,Zwickl2013,Lewandowski2015}.

Like modeling, troubleshooting is also a nonlinear, recursive process, though the goal is more narrow: to repair (or revise) a malfunctioning apparatus~\cite{Perez1991,Schaafstal2000,Jonassen2006}. Indeed, Zwickl \emph{et al.}~\cite{Zwickl2015} noted similarities between modeling and troubleshooting in their work characterizing student reasoning on an experimental optical physics activity. They found that, while troubleshooting, students sometimes engaged in ``what appeared to be a rapid modeling cycle involving a series of qualitative predictions and qualitative measurements \ldots in order to identify the source of the problem" (Ref.~\cite{Zwickl2015}, p.~8). It is precisely this overlap that we interrogate in the present work.

Troubleshooting is a task that spans a wide range of contexts, such as making medical diagnoses, maintaining manufacturing equipment, and debugging computer software systems (see Refs.~\cite{Perez1991,Schaafstal2000,Jonassen2006} for more comprehensive reviews of the troubleshooting literature). In the education research literature, major research foci include identifying the skills and knowledge required for effective troubleshooting~\cite{MacPherson1998,Johnson1988}, characterizing the novice-to-expert transition~\cite{Johnson1988,vanGog2005a,vanGog2005b}, and developing teaching strategies for troubleshooting~\cite{vanGog2006,vanGog2008,Kester2004,Kester2006,Johnson1999,Ross2007}. Of particular relevance to the present work, some previous work has focused on high school students' ability to troubleshoot simulated electric circuits~\cite{vanGog2005a,vanGog2005b,vanGog2006,vanGog2008,Kester2004,Kester2006}.

Three factors make electronics courses an ideal context for studying physics students' troubleshooting abilities. First, the physical systems and models with which students interact are relatively simple. Second, the electric circuits that students build during lab activities consist of low-cost, easy-to-replace components, thus facilitating multiple revisions to the experimental system. Finally, students often construct circuits that don't initially work, and the need to troubleshoot arises naturally in most lab activities. Previous work in the domain of electronics courses for physics students has focused on: characterizing college students' understanding of electric circuits~\cite{McDermott1992,*McDermott1993,Engelhardt2004,Coppens2012,Stetzer2013,Papanikolaou2015};
characterizing expertise-related differences among high school students troubleshooting simulated circuits~\cite{vanGog2005a,vanGog2005b}; and designing teaching strategies to develop college students' conceptual understanding~\cite{Shaffer1992,Getty2009,Mazzolini2011}, engage college students in model-based reasoning~\cite{Lewandowski2015}, and improve high school students' troubleshooting ability~\cite{vanGog2006,vanGog2008,Kester2004,Kester2006}. However, we are not aware of work that focuses on physics students' ability to troubleshoot physical (as opposed to simulated) electric circuits, or of work that focuses on the troubleshooting processes employed by post-secondary physics students.

In this paper, we report on a study that explores the overlap of modeling and troubleshooting in the context of an activity that is typical of an upper-division electronics course for physics students. In this study, eight pairs of students attempted to diagnose and repair a malfunctioning electric circuit. Using audiovisual data collected from two institutions, we characterize the cognitive tasks and model-based reasoning that students employed during this activity. The work herein builds on preliminary analyses of a subset of our data, which have been reported elsewhere~\cite{Dounas-Frazer2015,VanDeBogart2015}.

Our work expands current knowledge of instructional physics laboratory environments in two ways. First, we apply frameworks for both modeling and troubleshooting to a new domain in physics education, namely, upper-division electronics. Second, we examine the synergies of two nationally-recognized learning outcomes for lab courses, namely, (1) the ability to model experimental systems and (2) the ability to troubleshoot a malfunctioning apparatus. In doing so, we demonstrate that electronics courses---whose content is sometimes dismissed as not real physics---can engage students in important experimental physics practices.

This paper is organized as follows. In Sec.~\ref{sec:theory}, we describe the two theoretical perspectives that inform our work: a cognitive task analysis of troubleshooting and a framework for describing the modeling process. In Sec.~\ref{sec:context}, we provide institutional context for our study and a description of our study participants. In Sec.~\ref{sec:methods}, we describe our research methods, including a detailed description of the troubleshooting activity. Our results are presented in Sec.~\ref{sec:results} and discussed in Sec.~\ref{sec:discussion}. Finally, we summarize our findings and discuss future directions for our work in Sec.~\ref{sec:summary}.


\section{Theoretical perspectives}\label{sec:theory}

Throughout this work, we define troubleshooting as the process of repairing a malfunctioning system.  Troubleshooting is a type of problem-solving for which the solution state is known, but the troubleshooter must determine what information is needed for problem diagnosis~\cite{Jonassen2006}. Our goal is to identify and describe examples of how students use (or don't use) model-based reasoning while troubleshooting. We grounded our design and analysis in two different, complementary theoretical perspectives: a cognitive task analysis of troubleshooting~\cite{Schaafstal2000,Jonassen2006,Johnson1988}, and the Modeling Framework, which describes physicists' use of models when conducting physics experiments~\cite{Zwickl2015}. The motivations for using these two perspectives are twofold. First, we are able to map the Modeling Framework onto existing analyses of the troubleshooting process. Second, when analyzing students' approaches to troubleshooting, we use the cognitive troubleshooting tasks and the Modeling Framework to provide complementary coarse- and fine-grained descriptions of students' thought processes. In this section, we elaborate on each of these perspectives and identify areas of overlap. Because the system we consider here is a circuit, we provide examples from electronics to help clarify ideas throughout the discussion.

\subsection{Cognitive Task Analysis of Troubleshooting}

Cognitive task analysis is ``a family of methods used for studying and describing reasoning and knowledge" (Ref.~\cite{Crandall2006}, p.~3). Our summary of various cognitive task analyses of troubleshooting~\cite{Schaafstal2000,Jonassen2006,Johnson1988} explicates both the types of knowledge and the types of tasks that facilitate effective troubleshooting.

\subsubsection{Types of troubleshooting knowledge}

Other work~\cite{Jonassen2006,vanGog2005a} has identified six kinds of knowledge that facilitate competent troubleshooting: Domain, System, Procedural, Strategic, Metacognitive, and Experiential. \emph{Domain Knowledge} consists of the theories and principles upon which the system was designed~\cite{Jonassen2006}. In the case of a circuit, Domain Knowledge may include underlying principles like electron transport or conservation of charge as well as models like Ohm's Law or the Golden Rules for op-amps. These principles enable the troubleshooter to both represent the problem and identify relevant problem-solving operations~\cite{vanGog2005a}, such as which voltages to measure in order to determine whether or not a particular component is functioning properly.

\emph{System Knowledge} includes understanding of the structure and function of the system and the components within the system~\cite{Davis1983}. In a circuit, this may involve recognizing that a complex circuit is composed of multiple subsystems or identifying a particular resistor as a ``feedback resistor." System Knowledge further includes understanding of spatial representations of the system and flow control within the system~\cite{Jonassen2006}. For circuits, diagrams and schematics of the configuration of subsystems and components are common representations that are used to trace current through the system.

\emph{Procedural Knowledge} refers to the appropriate use of test equipment and procedures~\cite{Jonassen2006}. For electronic systems, this includes understanding how to use oscilloscopes, multimeters, and power supplies.

\emph{Strategic Knowledge} includes heuristic techniques and systematic approaches to troubleshooting the system~\cite{vanGog2005a}. Strategic Knowledge is an essential part of competent troubleshooting~\cite{Perez1991}. For example, experts employ particular sequences of operations to reduce the problem space by reducing the number of potential locations for faults~\cite{Johnson1988}. Three commonly used troubleshooting strategies are~\cite{Jonassen2006}:
\begin{enumerate}
\item Exhaustive, which involves identifying all possible faults and testing them one-by-one until the actual fault is discovered;
\item Topographic, which involves performing a series of tests that follow a trace through the system, either moving ``downstream" from a point where the system behaves correctly or ``upstream" from a point of malfunction; and,
\item Split-Half, which involves checking the functionality of the system at a midpoint in order to reduce the problem space by isolating the fault in one half or the other.
\end{enumerate}

\emph{Metacognitive Knowledge} ``is used to monitor [the troubleshooting process] by keeping track of the progress toward the goal state" (Ref.~\cite{vanGog2005a}, p.~237). Such monitoring is required in order to evaluate the effectiveness of a strategy and, if needed, to switch to a different strategy~\cite{Perez1991}. In ongoing work~\cite{VanDeBogart2015}, we are exploring the role of socially-mediated metacognition in troubleshooting~\cite{Goos2002}.

Finally, \emph{Experiential Knowledge} is the historical information accumulated by experienced troubleshooters~\cite{Jonassen2006}. Experiential Knowledge enables troubleshooters to propose likely faults by recalling historical information that links symptoms to likely causes. This process can be faster than relying on other types of knowledge to create logical connections between symptoms and causes based on the function of the system and its components. Recall of historical information has been shown to be a frequent diagnosis strategy among technicians in manufacturing~\cite{Bereiter1989}, machining~\cite{Konradt1995}, and maintenance~\cite{MacPherson1998} contexts. In an electronics course, neglecting to properly power a circuit is a common mistake. Thus, when students encounter a malfunctioning circuit whose output voltage is zero, students' experience might prompt them to immediately check power connections rather than speculate about the configuration or misbehavior of the circuit.

These six types of knowledge---Domain, System, Procedural, Strategic, Metacognitive, and Experiential---are brought to bear when attempting to repair a malfunctioning system, though they play different roles at different stages of the troubleshooting process.

\subsubsection{Cognitive troubleshooting tasks}\label{sec:CTAT}

\figtaskanalysis

Several models of the troubleshooting process exist, all of which describe the process as recursive and nonlinear~\cite{Jonassen2006,Johnson1988,Schaafstal2000}. In the present work, we draw on the cognitive task analysis proposed by Schaafstal \emph{et al.}~\cite{Schaafstal2000}, which subdivides troubleshooting into four iterative subtasks: Formulate Problem Description, Generate Causes, Test, and Repair and Evaluate. A graphical representation of these tasks is provided in Fig.~\ref{fig:taskanalysis}.

\emph{Formulate Problem Description} refers to the early stage of the troubleshooting process, during which the troubleshooter determines both what the system is doing wrong and what it is doing right~\cite{Schaafstal2000}. During this stage, the troubleshooter performs initial checks, measurements, and inspections of the apparatus. In an electronic system, this process involves orienting to the circuit~\cite{vanGog2005b} by building mental representations of the system structure and functions or using external representations~\cite{Jonassen2006}, such as schematics, datasheets, and equations. 

\emph{Generate Causes} involves generating causal hypotheses, either by recognition of common symptoms (typical of experts) or by using reasoning skills, functional thinking, and external documentation (typical of troubleshooters encountering a problem for the first time)~\cite{Schaafstal2000}. In addition to generating hypotheses that propose explanations for symptoms, this phase of troubleshooting also involves the use of Strategic Knowledge to propose procedures to facilitate identification of faults~\cite{Jonassen2006}. 

\emph{Test} involves performing measurements, tests, or checks to determine whether or not a proposed cause is indeed the actual fault that needs to be repaired. According to Shaafstal \emph{et al.}~\cite{Schaafstal2000}, this task includes ``choosing the right testing methods and the right testing [equipment]" as well as ``correctly setting up and operating the testing [equipment] and correctly reading the outcome of the test" (p.~79). In the case of a malfunctioning circuit, testing requires correct use of oscilloscopes, multimeters, and powers supplies. Performing tests further involves evaluating and interpreting the outcome~\cite{Schaafstal2000,vanGog2005b}, which requires that the troubleshooter form expectations about the behavior of a functional system and compare those expectations to the actual performance of the system. If the observed and expected outcomes of a test are in alignment, the proposed cause that informed the test must be rejected and the troubleshooter must generate additional causes. Alternatively, if a fault is identified, the next task is to repair the system.

Lastly, \emph{Repair and Evaluate} includes generating, enacting, and verifying solutions, in direct service to the goal of returning the system to its normal working state~\cite{Perez1991,Schaafstal2000,Jonassen2006}. Simple repairs involve replacing a component, though other types of repair are possible (e.g., soldering a broken connection). After performing a repair, evaluative measurements must be performed to determine whether the system is functioning normally. If not, the troubleshooter may conclude either that the repair did not address the fault or that the malfunction is due to multiple faults. In either case, the troubleshooter must return to the task of generating causes. If, on the other hand, the system behaves normally, then the troubleshooter may conclude that the repair is complete.

In this paper, our theoretical understanding of troubleshooting is partially informed by the six types of knowledge and the four subtasks described above. However, one goal of the present work is to understand the troubleshooting process through the lens of the Modeling Framework, which we describe in the following subsection.

\subsection{Modeling Framework}

\figmodelingframework

The Modeling Framework describes the dynamic process through which experimental physicists develop and refine models and apparatus. A diagram of the Modeling Framework is provided in Fig.~\ref{fig:modelingframework}. To explicate the framework, we define both ``models" and the process of ``modeling." In doing so, we draw heavily on the work of Zwickl \emph{et al.}~\cite{Zwickl2015}.

\subsubsection{Models}

Models are abstract representations of the real world. A well-defined model is associated with a target system or phenomenon of interest, and the model can be used for either explanatory and/or predictive purposes. Models are embedded in underlying principles and concepts relevant for understanding the target system. In addition, models are externally articulated through equations, diagrams, descriptions, and other representations. These representations are often informed by the topography of the target system and the flow of matter, energy, or information through the system. A circuit diagram, for instance, is a graphical representation of a circuit that shows how components are connected to one another and how charges flow through the circuit.

Importantly, models contain simplifying assumptions that yield tractable mathematical, graphical, and other representations. These assumptions limit the applicability of a model, meaning that users of the model must understand whether and when it can be accurately applied. Moreover, model limitations give rise to the possibility of model refinement by eliminating some assumptions, thus increasing the complexity of the model and broadening its scope of applicability. The iterative improvement of models to make them more accurate and sophisticated is one path in the process of modeling.

\subsubsection{Modeling}

Modeling is the process through which models and systems are brought into better agreement, either by refining the model or the target system itself. The Modeling Framework subdivides the target system into two parts, each with its own corresponding model (Fig.~\ref{fig:modelingframework}): the physical system and the measurement system. This subdivision reflects the fact that experimental physicists often operate measurement equipment in regimes where the limitations of that equipment become important. Such limitations must be accounted for either by making modifications to existing equipment or by developing an understanding of the tools' performance in new parameter regimes.

In many cases, the division between physical and measurement systems is fuzzy. For example, in  circuits such as the one considered here, the physical system consists of the circuit itself (wires, resistors, and other components) whereas the measurement system comprises voltmeters, ammeters, and other measurement tools. Whether power supplies and other ``test equipment" are included in the physical or measurement system reflects an arbitrary choice on the part of the modeler. Here, we include power supplies in the measurement system.

Modeling is a dynamic and iterative process, involving the following phases: model construction, interpretation, prediction, comparison, proposal, and revision. \emph{Model construction} refers to the development of models of the measurement and physical systems, depicted at the top left and right of Fig.~\ref{fig:modelingframework}, respectively. This process involves: identifying general principles and concepts that underly the model; making assumptions that simplify the model and identifying the corresponding limitations on the model's applicability; and choosing realistic values for model parameters.

While the model of the physical system is used to make \emph{predictions} about the performance of the physical apparatus, the model of the measurement system is used to \emph{interpret} raw data output by the measurement apparatus. In an optical system, this might involve using a known calibration factor to convert the output voltage of a photometer into a measurement of optical power. In electrical systems, when using a digital multimeter to measure the voltage of an oscillating signal, it is important to know whether the multimeter is displaying the amplitude of the signal or its root-mean-square value.

\emph{Comparison} is the act of comparing predictions to interpreted measurements. Discrepant measurements and predictions prompt physicists to \emph{propose} potential explanations for, and/or solutions to, those discrepancies. Resolving discrepancies requires a \emph{revision} to either the models or apparatus. The framework describes four pathways of revision: refine the measurement system model, the measurement system apparatus, the physical system apparatus, or the physical system model. Prioritization of one particular revision pathway over others depends on many factors, including the nature of the task. For example, based on the definition of troubleshooting used here, a troubleshooting activity will likely result in revision to the physical system apparatus.
 
\subsection{Synthesizing the Frameworks}
While the cognitive task analysis of troubleshooting and the Modeling Framework provide two distinct perspectives through which to understand the troubleshooting process, they are nevertheless connected. In this subsection, we synthesize these two perspectives by describing how both the types of knowledge and the cognitive tasks involved in troubleshooting relate to the modeling process.

\subsubsection{Modeling and types of troubleshooting knowledge}

Domain, System, and Procedural Knowledge can be directly connected to modeling. For example, these types of knowledge are required for the construction of models of the physical and measurement systems. Strategic and Metacognitive Knowledge, on the other hand, are only implicitly connected to modeling. For example, the process of modeling involves deciding which measurements to perform, in what order, and for what purpose. Alternatively, in response to a discrepancy between measurement and prediction, a physicist must decide which of the four revision pathways to enact. While such strategic and metacognitive decisions are necessary parts of the modeling process, they are not explicitly represented in the Modeling Framework. Rather, they are implicitly embedded in the arrows of Fig.~\ref{fig:modelingframework}: each arrow represents different possible metacognitive and strategic choices on the part of the experimentalist while navigating between different phases of the modeling process.

Depending on the circumstances, Experiential Knowledge can also be implicitly embedded in the Modeling Framework. For example, a troubleshooter may rely on historical information when making decisions about what to measure or what to revise. In this sense, the role of Experiential Knowledge in modeling is similar to the roles of Strategic and Metacognitive Knowledge. In other cases, however, Experiential Knowledge may limit the relevance of the framework for understanding a particular instance of troubleshooting. Experienced troubleshooters call on event schemas based on their historical experience with a system and its specific fault tendencies, often shortening their diagnostic process~\cite{Jonassen2006}. Using schemas to solve problems quickly is a common feature of expert problem-solving in physics and other contexts~\cite{Redish2002}. Thus, Experiential Knowledge may facilitate direct connections between symptoms and diagnoses without the need to engage in the recursive, nonlinear processes the Modeling Framework was designed to describe. In these situations, the framework may not be the most appropriate tool for characterizing the troubleshooting process.

\subsubsection{Modeling and cognitive troubleshooting tasks}

The cognitive troubleshooting tasks provide a taxonomy for some of the modeling phases. For example, consider the role of measurement in troubleshooting. Measurements can be classified into three types according to the cognitive task with which they are affiliated: formative measurements, which serve to formulate the problem description during initial stages of the troubleshooting process;  diagnostic measurements, used to test causal hypotheses during the testing phase; and evaluative measurements, used to determine whether the system has been restored to its functional state after a revision has been made. Similarly, the cognitive troubleshooting tasks discriminate between two types of proposals: proposed explanations for discrepancies between measurement and prediction, which facilitate generation of causes; and proposed solutions for resolving those discrepancies, which inform repairs to the system.

Conversely, the Modeling Framework provides a taxonomy of repair types: any of the four revision pathways in the Modeling Framework could constitute a type of repair during the troubleshooting process. In our study, however, repairs primarily consisted of revisions to the physical system apparatus, ultimately privileging one recursive pathway in the Modeling Framework (shaded in dark grey in Fig.~\ref{fig:modelingframework}). 

One major goal of the present work is to identify and describe examples of how students use (or don't use) model-based reasoning while troubleshooting. To help us unpack the mapping between the Modeling Framework and the troubleshooting process, we designed an observational study in which pairs of students were tasked with repairing a malfunctioning circuit. In the following section, we describe the institutional context and the participants involved in our study.


\section{Context and participants}\label{sec:context}

\demotable

Our study was carried out at two universities, the University of Colorado Boulder (CU) and the University of Maine (UM). Both institutions are predominantly white four-year public research universities with high undergraduate enrollment. CU is a large, more selective institution with very high research activity; UM is a medium, selective institution with high research activity~\cite{Carnegie2011}. Demographic information about the physics programs at each institution is summarized in Table~\ref{tab:demographics}. These demographics reflect the makeup of the students enrolled in the Electronics Courses at CU and UM as well as those who participated in our study.

The Electronics Courses at CU and UM share many similarities. Each course is required for all physics majors, with students typically completing the course during their third year of instruction. Both courses convene three times per week: twice for one-hour lectures and once for a multi-hour lab (three hours at CU, two at UM).  Both Electronics Courses consist of 2--3 lab sections, with 15--20 students per section at CU and 5--8 at UM. Lectures and labs are taught by tenured or tenure-track physics faculty members. Teaching and/or learning assistants support instruction at each institution. Both courses focus on analog components (e.g., op-amps, diodes, and transistors) and circuits (e.g., dividers, filters, and amplifiers). To learn this material, students work in pairs on guided lab activities. There is no formal instruction about troubleshooting in either course; instead, discussion about troubleshooting is limited to impromptu conversations between students and instructors in response to problems that inevitably arise during lab.

\figdesign

The CU and UM courses differ in several ways. At CU, for example, the course is offered every semester and enrollment varies from 30--60 students per term. Students have keycard access to the lab room at all hours of the day, including weekends. In addition, the CU course culminates in a five-week final project. Finally, the CU Electronics Course was recently redesigned to engage students in modeling of canonical measurement equipment and analog circuits~\cite{Lewandowski2015}, in alignment with consensus learning goals for upper-division labs identified by physics faculty members at CU~\cite{Zwickl2012}. Additional learning goals for this course were identified through interviews with graduate students who use electronics as part of their experimental physics research~\cite{Pollard2014}. 

At UM, on the other hand, the course is offered only in the fall, with roughly 10--15 students per term. Moreover, the UM Electronics Course is designated a ``writing intensive" laboratory course, which means that students are required to complete formal lab write-ups that are critiqued by an outside technical writing expert (in addition to the Electronics instructor).

Study participants were physics or engineering physics majors enrolled in the Electronics Course at either CU or UM during Fall 2014. During that time, two of the authors (HJL and MRS) taught lab and lecture sections for the CU and UM courses, and one of the authors (KLVDB) was a teaching assistant for the UM course. We solicited participation in the study via email and in-person requests during the last few weeks of Fall 2014 and the first few weeks of Spring 2015. The study was not an official part of either the CU or UM course, and no course credit was associated with participation in the study. Participants were consenting volunteers who received small monetary incentives for their participation.

In total, 16 students participated in the study, 8 each from CU and UM. We interviewed students in pairs, forming four pairs at each institution. Two pairs consisted of students who were lab partners during their Electronics Course, and six pairs consisted of students who were not lab partners. The latter six pairs were formed by the research team by pairing students who had expressed interest in participating in the study. Fifteen participants earned grades ranging from A to B-- in their Electronics Course, which required students to work with the components and systems used in our study. One student did not receive a passing grade due to a failure to submit all of the lab reports and lab notebooks, per the grading policy of the course. During the interviews, all eight student pairs attempted to repair a malfunctioning electrical circuit, as described in the following section.


\section{Methods}\label{sec:methods}

To probe whether and how students engaged in model-based reasoning while troubleshooting, we conducted Think-Aloud Pair Problem Solving (TAPPS) interviews with eight pairs of students. TAPPS interviews involve students working on an activity while concurrently verbalizing their thoughts aloud~\cite{Ericsson1993,vanSomeren1994}, providing the research team with information on student reasoning about actions and outcomes~\cite{Taylor2000}. In the troubleshooting literature, TAPPS interviews have been used in both training~\cite{Johnson1999} and research~\cite{Pate2011} contexts. During our TAPPS interviews, student pairs were tasked with repairing a malfunctioning electrical circuit, namely, an inverting cascade amplifier. In this section, we describe the design of the TAPPS interview and elaborate on data collection and analysis methods.

\subsection{Troubleshooting activity}\label{sec:task}

We designed an inverting cascade amplifier that contained two subsystems, or stages: a noninverting amplifier (Stage~1) and an inverting amplifier (Stage~2). Each stage consisted of an op-amp and two resistors: $R_1$ and $R_2$ in Stage 1, and $R_3$ and $R_4$ in Stage 2. A diagram of the circuit is provided in the left panel of Fig.~\ref{fig:design}.

\subsubsection{Functioning circuit behavior}

In a functional circuit, each stage would have amplified its input voltage by a multiplicative factor called the gain. The theoretical gains for Stages~1 and 2 were $G_1=(1+R_2/R_1)$ and $G_2=-R_4/R_3$, respectively.  Because the two stages were connected in series, the overall gain of the cascade amplifier was the product of the gains of Stages 1 and 2: $G_{\mathrm{tot}}= G_1G_2$. For a given input voltage, $\Vin$, a functional circuit's output voltage, $\Vout$, would be given by:
\begin{equation}\label{eq:transfer}
\Vout=-\Vin(1+R_2/R_1)(R_4/R_3).
\end{equation}
Equation~\ref{eq:transfer} is valid under the following two conditions: first, the magnitude of the input voltage is sufficiently small that the outputs of Stages~1 and 2 are smaller in magnitude than the corresponding supply voltages, $V_{\mathrm{S}1\pm}$ and $V_{\mathrm{S}2\pm}$; and second, the frequency of the input is sufficiently low that bandwidth limitations of the op-amps can be ignored.

The resistances and voltages characteristic of a functioning circuit are given in Table~\ref{tab:faults}. Nominal resistances and supply voltages were communicated to interview participants via the schematic and datasheet which were provided to them during the activity. For these nominal values, $G_1=2$, $G_2=-10$, and $G_{\mathrm{tot}}= -20$. That is, in a functional circuit, Stage~1 would double the input voltage, Stage~2 would both invert the output of Stage~1 and amplify it by a factor of 10, and the circuit as a whole would invert the input voltage and amplify it by a factor of 20. In the context of alternating current (ac) signals, ``inverting" is equivalent to shifting the phase of the signal by 180$^\circ$. Plots~(a) and (b) in the right panel of Fig.~\ref{fig:design} show the outputs of Stages~1 and 2 when an ac signal is input to a functioning inverting cascade amplifier.

Finally, each of the op-amps in a functional circuit would obey the following ``Golden Rule" for op-amps in a closed loop with negative feedback:  there is zero voltage difference between the two input terminals.

\tabfaults

\subsubsection{Malfunctioning circuit behavior}

To ensure that students engaged in more than one iteration of troubleshooting, we introduced two faults in the circuit, as shown in the left panel of Fig.~\ref{fig:design}. First, the resistor $R_3$ had a value of 100~\ohm\ rather than the nominal value of 1~k\ohm, increasing the actual gain of Stage~2 by an order of magnitude compared to the nominal gain. Second, we used a malfunctioning op-amp in Stage~2, which manifested in a direct current (dc) output voltage of $-15$~V regardless of the input voltage. Both faults were localized in Stage~2 so that the cascade amplifier consisted of both a functional subsystem (Stage~1) and a malfunctioning one (Stage~2), making it possible for students to use the Split-Half Strategy. The output of the malfunctioning circuit is shown in Plot~(c) of Fig.~\ref{fig:design}. Additional characteristics of the malfunctioning circuit are provided in Table~\ref{tab:faults}.

If students were to replace the 100~\ohm\ resistor with a 1~k\ohm\ resistor but leave the faulty op-amp in the circuit, the output of the circuit would remain unchanged. On the other hand, if student were to replace the faulty op-amp with a functioning chip but leave the 100~\ohm\ resistor in the circuit, the circuit would effectively be a functioning inverting cascade amplifier with an overall gain of $-200$. In this case, the output of the circuit could potentially be a clipped signal. A``clipped signal" refers to a sinusoid with flattened peaks, as shown in Plot~(d) of Fig.~\ref{fig:design}. The flattening is due to limitations of the op-amp, which cannot output voltages larger than about 13~V in magnitude. A gain of $-200$ would result in clipping for any ac input signals with amplitude larger than only about 6~mV. Even in a functioning circuit, clipping may arise for input ac signals with amplitude larger than about 650~mV.

In the malfunctioning state, the op-amp in Stage~2 did not obey the Golden Rule for op-amps. There was a nonzero dc voltage difference between the input terminals of the faulty op-amp, as indicated in Table~\ref{tab:faults}.

To troubleshoot the circuit, student pairs had access to test and measurement equipment that was typical of the equipment used in their Electronics Course. All students had access to an oscilloscope, digital multimeter, low-voltage dc power supply, ac function generator, pliers, wire strippers, various types of cables, and extra resistors, op-amps, and wire. CU students used a breadboard that needed to be connected to external voltage sources. UM students, on the other hand, used a commercial prototyping board that had on-board ac and dc voltage sources. In both cases, the malfunctioning circuit was pre-built on the board by the research team. A photograph of the setup used at CU is provided in Fig.~\ref{fig:photo}.

We designed the think-aloud troubleshooting task to closely mimic the types of troubleshooting events that students typically encounter during the Electronics Course. After students completed the task, we asked them to comment on the extent to which the task felt like a typical Electronics Course activity to gain insight into the ecological validity of the task~\cite{Bronfenbrenner1977}.  Student responses indicated that the think-aloud activity was similar to their typical course experiences along several dimensions, including the components used, the equipment used, the faults they encountered, and the processes they used to troubleshoot the circuit. We conclude that the activity was similar to students' in-class experiences.

\subsection{Data collection}

We conducted TAPPS interviews in which pairs of students were tasked with diagnosing and repairing the malfunctioning inverting cascade amplifier shown in Fig.~\ref{fig:design}. At the start of each task, the interviewer provided students with the following materials: a schematic diagram of the circuit, a datasheet for the op-amp, and a pre-built malfunctioning circuit. The circuit schematic and corresponding text are shown in Fig.~\ref{fig:schematic}. No expressions for the gains of either of the two individual stages were provided to the students.

\figphoto

The interviewer read a short prompt to the students before they began the task. The prompt framed the activity as follows:
\begin{quote}
For this activity, you will be repairing a malfunctioning circuit. Specifically, you'll be working with an inverting cascade amplifier, described on this page here [Fig.~\ref{fig:schematic}]. For context, let's imagine that some of your peers built this circuit as part of class. They built the circuit using the same chip you've been using in class this semester. Here's the standard data sheet for that chip. Your tasks are to diagnose any issues and make the circuit work properly.

This interview is very similar to what you've been doing in class. You'll have access to much of the equipment from class, including power supplies, measurement tools, and a limited selection of electrical components. One difference from class is that you're working with a circuit someone else built. Another difference is that I'm interested in what you say to yourself as you perform this task, so I will ask you to talk aloud as you work on the circuit.

What I mean by talk aloud is that I want you to say out loud everything that comes into your mind while doing the task. Put another way, I want you say out loud what you might otherwise say to yourself silently. Of course, you should also feel free to ask each other questions and interact as you would when working together in [the Electronics Course]. But the more you both say out loud what you're thinking in your head, the more helpful it will be.

Act as if I am not in the room. Just keep talking. If you are silent for any length of time, I will remind you to keep talking aloud.
\end{quote}

After reading the prompt, the interviewer asked the students to begin working on the task. During this time, the interviewer interacted only minimally with the students. The activity ended when either the students repaired the circuit or an hour had passed. After the activity was over, the interviewer asked students a few short follow-up questions, including a question about the extent to which the activity felt typical of students' experiences in their Electronics Course.

The interviewers' prompts and follow-up questions accounted for only 5--10 minutes per activity. In six interviews, students spent 40--45~min troubleshooting the circuit. In the other two interviews, students repaired the circuit in 20--25~min. Together, all eight TAPPS interviews lasted a total of six hours, with about five hours devoted to troubleshooting the circuit. Video and audio data were collected for all interviews, and audio data were fully transcribed.

\figschematic

\subsection{Data analysis}
Our approach to data analysis involved two parts. First, we used the cognitive troubleshooting tasks to code successive two-minute intervals that spanned the duration of each of the interviews. Second, we used the Modeling Framework to code two types of events present in most interviews: (1) isolation of the second stage as the source of faults, and (2) evaluation of the circuit after replacing the faulty op-amp.

Both the cognitive troubleshooting tasks and the Modeling Framework were used as \emph{a priori} analysis schemes. For each scheme, we initially developed operational code definitions based on global definitions from the troubleshooting and modeling literature~\cite{Schaafstal2000,Zwickl2015} and a review of the content logs. Operational definitions were refined through iterative cycles of collaborative coding and discussions with the research team. Whereas codes related to the Cognitive Troubleshooting Tasks were applied to a total of 5 hours of troubleshooting activity across all 8 interviews, the Modeling Framework codes were applied only to a total of about 30~minutes across all eight TAPPS interviews. Below, we describe the coding schemes in greater detail.

\subsubsection{Cognitive troubleshooting tasks}\label{sec:CTT}

To characterize students' approach to troubleshooting the malfunctioning cascade amplifier, we developed codes corresponding to the four cognitive troubleshooting tasks described in Sec.~\ref{sec:CTAT}: Formulate Problem Description, Generate Causes, Test, and Repair and Evaluate. Each code was associated with three or four subcodes that were generated through an emergent and iterative process, starting with a review of the audiovisual recordings. The subcodes are listed in Table~\ref{tab:tasksubcodes}. To apply the subcodes to our data, we divided each video into successive two-minute intervals. We collaboratively coded each interval through three iterations of coding, discussion, and refinement of subcode definitions and applications. Depending on the nature of student activity during a given time interval, we assigned no subcodes, one subcode, or multiple subcodes to the interval. 

As an example of our coding scheme, the Formulate Problem Description subcodes (italicized font) and their operational definitions (normal font) were:
\begin{itemize}
\item \emph{Map circuit onto schematic and/or datasheet:} Students orient themselves to the circuit topographically by mapping the circuit onto the schematic or datasheet. This typically involves looking back and forth between the circuit, schematic, and datasheet, articulating which chip corresponds to which stage, which resistors correspond to $R_1$--$R_4$, which pins are input and output, and/or where the power and ground connections are located.
\item \emph{Discern functions of systems, components:} Students do at least one of the following: identify the circuit or one of its subsystems as an inverting or noninverting amplifier; discuss the function of a component (e.g., ``this is a feedback resistor"); or rationalize the absence of capacitors (e.g., ``capacitors are just needed to clean up high-frequency noise from the signal").
\item \emph{Perform formative measurements:} Students perform initial checks of the circuit configuration, resistor values, pin voltages, or the performance of the test and measurement equipment. These measurements are typically accompanied by statements like, ``I'm just trying to figure out what's going on."
\end{itemize}

According to our subcode definitions, a measurement of, say, voltage could be an example of either Formulate Problem Description, Test, or Repair and Evaluate depending on whether it was performed in a formative, diagnostic, or evaluative capacity. For example, an initial check that the op-amps are powered would be an example of a formative measurement. On the other hand, measuring the midpoint voltage as part of a Split-Half Strategy would constitute a diagnostic measurement. Finally, checking the output signal after replacing the op-amp in Stage 2 would be an evaluative measurement.

Although Schaafstal \emph{et al.}~\cite{Schaafstal2000} include setup and operation of test and measurement equipment in their global definition of ``performing the test", we did not include these actions in our final definition of Test. In our dataset, students were adjusting settings on the oscilloscope, multimeter, power supply, and/or function generator throughout the activity, which effectively contributed a ``constant background" of this aspect of testing to our analysis.

\subsubsection{Modeling Framework}\label{sec:MF}
To characterize students' model-based reasoning during the troubleshooting process, we developed codes based on the Physical System half of the Modeling Framework. We applied these codes to two types of events: (1) isolation of the second stage as the source of faults, and (2) evaluation of the circuit after replacing the faulty op-amp. For both types of events, we used the Modeling Framework codes to perform line-by-line analyses of the corresponding transcribed student dialogue. A detailed example of this approach is provided elsewhere~\cite{Dounas-Frazer2015}.

\taskcodes

In total, we identified 12 excerpts (5 isolation-type events and 7 evaluation-type events) which lasted about 2--3 minutes each. Excerpts for isolation-type events started when one or both students suggested measuring the output of Stage~1 and ended when the students concluded that Stage~1 was functioning as expected and the faults could therefore be isolated in Stage~2. Excerpts for evaluation-type events started just after the students replaced the faulty op-amp and ended once the students concluded that the circuit as a whole had been repaired and was now behaving as expected. Isolation-type events, when they occurred, took place between half and two-thirds of the way though the think-aloud activity. Evaluation-type events, on the other hand, spanned the last three minutes of the activity.

The Modeling Framework codes (bolded font) and their operational definitions (normal font) were:
\begin{itemize}
\item {\bf Model Construction (Physical System)}: Students do any of the following: model the circuit as being comprised of two abstract subsystems, each with its own gain; identify relevant principles and concepts from electronics, such as the Golden Rule for op-amps in a closed loop; or identify limitations of the transfer function, such as the gain-bandwidth product or voltage-related limits on the amplitude of the output signal.
\item {\bf Prediction}: Students compute expected outputs, such as: the phase, amplitude, or frequency of the signal at various points in the circuit; or the gain of a subsystem or the system as a whole. This includes articulating expectations about what would happen if a component had a different value (e.g., they compute the gain of a hypothetical circuit with, say, $R_4=1\;\mathrm{k}\Omega$).
\item {\bf Comparison}: Students compare expected and measured values of amplitude, phase, or frequency of a signal (e.g., ``We see 100~mV, but it should be 4~V."). This includes making relational statements about the size of a signal (e.g., ``This signal is too small.") and making evaluative judgements about the observed signal (e.g., ``This is strange," or ``This isn't what it's supposed to do.").
\item {\bf Proposal}: Students suggest explanations for a discrepancy between measurement and prediction. Alternatively, students suggest solutions for bringing the actual performance of the circuit into alignment with expectations.
\item {\bf Revision (Physical System)}: Students change the circuit configuration, replace a resistor, or replace an op-amp in response to a discrepancy between measurement and prediction.
\end{itemize}
The definition of Prediction does not require that students' computations or expectations are correct. Neither does the definition require that computations or expectations are carried out or articulated before a measurement is performed. Indeed, in our dataset, many instances of Prediction occur \emph{after} a measurement takes place in an effort to determine whether or not the measured value ``makes sense."

While Zwickl \emph{et al.}~\cite{Zwickl2015} include ``identify parameters" as part of constructing models, we did not address this aspect in our final definition of Model Construction. In the episodes we chose to analyze with the Modeling Framework, there were no instances of students identifying parameters in service of constructing a model of the circuit.


\section{Results}\label{sec:results}

\taskresults

We describe the troubleshooting processes of eight pairs of students. Since we are not performing a comparative analysis, we do not distinguish between students at CU and those at UM. Pairs of students are labeled A--H and individual students are labeled according to their pair membership. For example, the students in Pair~A are labeled A1 and A2. When providing examples of students' verbalizations, we indicate the speaker as well as the time interval during which they were speaking.

In the following subsections, we describe the students' troubleshooting process and the changes they made to the circuit. Using results from two coding schemes (detailed in Secs.~\ref{sec:CTT} and \ref{sec:MF}), we show that each pair of students engaged in all of the cognitive troubleshooting tasks and demonstrated model-based reasoning during strategic and/or evaluative stages of the troubleshooting process.

\subsection{Cognitive tasks}

The results of coding for the four cognitive troubleshooting tasks are shown in Fig.~\ref{fig:tasks}. Formulate Problem Description, Generate Causes, Test, and Repair and Evaluate are represented as orange, blue, green, and yellow bands, respectively. Based on these codes, several patterns can be discerned. For example, all eight pairs engaged in all four cognitive troubleshooting tasks. Most pairs transitioned from formulating the problem description to generating causes about halfway through the activity, though the transition isn't always clear cut. Testing happened almost continuously throughout the duration of activity, whereas repairs and evaluations were performed more sporadically.

The filled black triangles and stars in Fig.~\ref{fig:tasks} correspond to times when students replaced the 100~\ohm\ resistor and the faulty op-amp, respectively. These symbols reveal additional patterns. For example, all pairs replaced the resistor and all-but-one replaced the op-amp. In each case, the resistor was replaced before the op-amp was replaced. Indeed, most pairs replaced the resistor early in the troubleshooting process, while they were still formulating the problem description and before they started generating causes. In addition to replacing the 100~\ohm\ resistor and/or the faulty op-amp, many pairs performed additional, unnecessary revisions to the circuit. In Fig.~\ref{fig:tasks}, such revisions sometimes manifest as yellow blocks which contain neither a triangle nor a star.

The filled black circles in Fig.~\ref{fig:tasks} indicate the times when students employed the Split-Half Strategy. Pairs~D--H employed this strategy at some point during the second half of the activity, often within a few minutes after the onset of generating causes. Pairs who used the Split-Half Strategy did not repair the circuit significantly faster or slower than those who did not. For example, Pairs~A and E repaired the circuit about twice as quickly as other groups, but only E employed the Split-Half Strategy.

Below, we elaborate on the results from our cognitive task coding scheme.

\subsubsection{Formulate Problem Description}

Formulate Problem Description refers to the early stage of the troubleshooting process, during which students oriented themselves to the activity and performed formative measurements to determine what was working and what was not. Students engaged in this task throughout the first half of the activity. We report verbalizations that reflect three different aspects of problem formulation: (1) mapping the circuit to the schematic, (2) discerning the intended function of the circuit, and (3) performing initial measurements and inspections to check various aspects of the circuit.

All eight pairs mapped the schematic to the circuit almost immediately after the interviewer finished giving instructions. For example, as soon as the activity started, Student~B1 said:
\QUOTE{So, we should identify which one [stage] is which to start with. \ldots\ This is a noninverting [stage] to start with. This part only. And the next one is an inverting [stage].}{B1}{2:35}{3:09}
Here B1 correctly mapped the components in the circuit to their symbolic representations in the schematic. In addition, B1 simultaneously recognized the existence of noninverting and inverting stages, an important part of discerning the intended function of the circuit and its subsystems.

Seven pairs (B--H) identified the existence of the two stages and discerned their intended functions. For example, soon after examining the schematic for the first time, Student~E1 said:
\QUOTE{That makes sense, just like inverting and non-inverting smashed together.}{E1}{2:45}{2:49}
Thus, E1 parsed the circuit as comprising two subsystems early in the activity, \emph{before} performing any measurements. Pairs B and C did the same. Pairs~D and F--H, on the other hand, discussed the intended function of the circuit \emph{after} they had begun performing measurements and generating hypotheses about potential faults. In these cases, discerning the function of subsystems was a crucial part of understanding the results of diagnostic measurements. For example, after measuring the output of Stage~1, Student~F2 said:
\QUOTE{And it [the output of Stage~1] shouldn't be inver--, should be \ldots\ An inverting amplifier is connected to the,  $\Vin$ is connected to the negative terminal, right? \ldots\  So it [the output of Stage~1] shouldn't be inverted. \ldots\ The second one [stage] is inverting.}{F2}{24:52}{25:27}
Here F2 discerned the function of the subsystems in response to a diagnostic measurement.
In Fig.~\ref{fig:tasks}, this and similar occurrences in Pairs~D, G, and H are depicted by instances of Formulate Problem Description (orange bands) that occur after the onset of Generating Causes (blue bands).

To formulate the problem description, students performed formative checks of configuration, resistances, and voltages. All eight pairs began checking the circuit configuration within a few minutes of receiving instructions. For example, after mapping the circuit to the schematic, Student~A1 said:
\QUOTE{Let's see if everything's connected right first off.}{A2}{2:50}{2:53}
Here A1 suggested checking the circuit configuration as one of the first steps in the troubleshooting process. Six pairs (B--E, G, and H) also measured the resistances of all four resistors during this phase, leading them to identify and replace the 100~\ohm\ resistor early in the activity. Students employed the Topographic Strategy when checking circuit configuration and resistor values, starting from the input and tracing through the circuit to the output, or vice versa. For example, when deciding to measure the resistor values, Student~G1 said:
\QUOTE{Check all the resistor values and then make sure they're all connected. \ldots\ You want to just start at the bottom [input] and go to the top [output]?}{G1}{2:32}{2:50}
Here G1 suggested checking the resistors in order of where they occur along the path from the input to the output of the circuit. Each pair also performed formative voltage measurements to ensure that the circuit was properly powered and grounded. In all cases, measurement of the output signal triggered students to begin generating causes and testing the circuit.

\subsubsection{Generate Causes}

Generate Causes involves making hypotheses about potential faults. Students generated causes throughout the second half of the activity, starting after measuring the output signal for the first time. Students proposed many different potential explanations for the malfunctioning behavior of the circuit, including short circuits, saturation, and faulty chips. We provide three examples of proposals: one that was dismissed, one that resulted in a revision to the circuit, and a third proposal that gave rise to diagnostic measurements.

The following explanation, proposed by Student~C1, is an example of a hypothesis that was dismissed:
\QUOTE{It's [the circuit is] probably saturated.}{C1}{25:44}{25:48}
Here, C1 suggested that the observed dc output signal was caused by output limitations of the op-amps; when the expected output voltage exceeds the limitations of the op-amp, the circuit is sometimes referred to as being ``saturated." This idea was immediately dismissed by C2, who noted that the output was a dc signal rather than the clipped ac signal characteristic of saturated amplifier circuits, as in Plot~(d) of Fig.~\ref{fig:design}.

Not all hypotheses were dismissed. For example, during a brainstorming session, Student~F1 simultaneously proposed an explanation and a revision:
\QUOTE{Could the op-amps be faulty? Should we just replace them with new ones? For the second op-amp, let's just replace it with a new one.}{F1}{40:18}{40:26}
Here F1 suggested that faulty op-amps might be the cause of the dc output signal, and then immediately proposed replacing the op-amp in Stage 2 with a new chip. F1 and F2 went on to replace the op-amp in Stage~2 as suggested.

Many proposed causes informed follow-up diagnostic measurements. For instance, upon seeing a dc output signal equal to the negative supply voltage, Student~B1 said:
\QUOTE{Maybe this red one [wire], the power is somehow touching the output.}{B1}{27:57}{28:03}
Here, B1 suggested there might be a short circuit connecting the negative dc supply voltage and the output of the circuit. After making this suggestion, B1 went on to perform a diagnostic visual inspection of the circuit for such a short.

\subsubsection{Test}

Test involves performing diagnostic measurements to determine whether or not a \emph{proposed} fault is an \emph{actual} fault. Test further includes prioritizing measurements, making plans, and making predictions about expected outcomes. Students engaged in testing throughout the duration of the think-aloud activity, focusing on prioritizing, planning, and predicting during the first half of the activity. During the second half of the activity, students began performing diagnostic measurements to check proposed causes. We report verbalizations that reflect the planning, prioritizing, and predicting aspects of testing.

The following exchange is an example of students prioritizing measurements. Immediately after observing the erroneous dc output of the circuit for the first time, Pair~E discussed their plans for diagnostic measurements. Student~E1 outlined the following plan:
\QUOTE{What we could do is get out a probe and we can just go through the first one [stage] and measure $\Vout$, and we could see if that's what we expect it to be.}{E1}{15:41}{15:52}
Here, E1 suggested performing diagnostic voltage measurements at various points in the first stage as well as of the output. E2 agreed with this plan, and suggested performing diagnostic measurements of the supply voltages as a follow-up:
\QUOTE{Yeah, for sure. And then we'll measure all the power, too, and make sure it's doing what it should be doing.}{E2}{15:52}{16:00}
In this exchange, making measurements of Stage~1 was prioritized over checking supply voltages.

Near the end of their troubleshooting process, Students~H1 and H2 came up with the following plan to determine whether the op-amp in Stage~2 was faulty:
\QUOTE{It seems like it [the cause of the malfunctioning behavior] might be that second op-amp.  So let's just put the input [signal] to the---let's pull---yeah, let's just hook the input [signal] to the 1k [resistor] and look at the output [of Stage 2]. \ldots\ We should get inverted times ten output.}{H2}{42:15}{43:03}
Here H2 recommended connecting the input signal, $\Vin$, to the input of Stage~2, thus bypassing Stage~1 altogether. This plan, which the students carried out, allowed the students to test Stage~2 as an isolated system. In addition to making a plan, H2 also made a new prediction; H2 correctly predicated that the second stage, if functioning properly, would invert the input signal and amplify it by a factor of 10.

Through testing, students were able to determine that some of their proposed causes were indeed actual faults in the circuit. For example, after isolating the second stage to test the performance of the second op-amp, Pair~H concluded that the second op-amp was indeed faulty and in need of replacement. Testing thus paved the way for repairs to the circuit.

\subsubsection{Repair and Evaluate}

Repair and evaluation involves proposing, enacting, and evaluating revisions to the circuit. Repairs and evaluation typically happened in short bursts (Fig.~\ref{fig:tasks}). Here we focus on three aspects of repairing and evaluating the circuit: replacing the 100~\ohm\ resistor, making erroneous revisions to the circuit, and replacing the faulty op-amp.

All eight pairs correctly identified the 100~\ohm\ resistor as a fault and replaced it with a 1~k\ohm\ resistor. Six pairs (B--E, G, and H) identified the resistor early in the activity, while checking that the circuit was constructed properly. In these cases, there could be no evaluative measurement of the revised circuit's performance because the students hadn't yet observed the output signal. Pairs A and F, on the other hand, identified the resistor as part of the testing process. Both of these pairs performed a quick evaluative measurement of the output voltage to determine whether their revision repaired the circuit. For example, after replacing the 100~\ohm\ resistor with a 1~k\ohm\ resistor, Student F2 measured the output signal and said:
\QUOTE{That resistor was wrong. \ldots\ But it's [the circuit is] still not working.}{F2}{30:26}{30:32}
F2's statement that the circuit was ``still not working" frames the measurement of the output signal as an evaluative measurement.

Five pairs made unnecessary or erroneous changes to the circuit. There were three types of erroneous revisions. First, Pair~A replaced $R_1$, nominally 460~\ohm, with a resistor whose measured resistance was closer to the nominal value than the original resistor. Second, Pairs~B, D, and H changed the circuit configuration. In each case, these changes were due to incorrect mapping of the circuit to the schematic or datasheet and the students eventually realized their mistake and restored the original circuit configuration. Finally, Pairs A--C replaced the op-amp in Stage~1 with a new chip.

For Pairs~A and B, the decision to replace the first op-amp was based on an Exhaustive Strategy in which both op-amps were replaced simultaneously. For example, after Pair~A had checked the circuit configuration, measured the resistor values, and replaced $R_3$, Student~A2 made the following suggestion:
\QUOTE{Want to replace the op-amps? \ldots\ We've gone over all the wiring twice at this point. We've gone over each individual resistor. So, do you have spare op-amps here?}{A2}{19:46}{20:07}
A2 reasoned that, since the wiring and resistors had previously been checked, the next step was to replace the op-amps---the only remaining type of component that had not yet been tested. The suggestion to replace the op-amps was the result of having exhausted other potential causes for the malfunctioning behavior of the circuit.

Pair~C, on the other hand, replaced the first op-amp individually. After observing noisy voltage measurements at various points along the circuit, C2 offered the following explanation:
\QUOTE{You know, [another student] and I had this problem the other day and the problem was the op-amp was just bad. \ldots\ We popped in a new op-amp and it fixed it.}{C2}{26:41}{26:49}
C1 responded by suggesting that they ``swap out that first op-amp." Here Pair~C relied on Experiential Knowledge to propose a solution to the noisy voltage measurements: C2 had previously encountered noise in an op-amp circuit, the cause of the noise was a ``bad" op-amp, and the solution path was to replace the op-amp with a new chip.

Seven pairs (A--C and E--H) correctly identified the op-amp in Stage~2 as a fault and replaced it with a functional chip, successfully repairing the malfunctioning cascade amplifier. Each of these successful pairs replaced the resistor before replacing the op-amp. Upon repairing the circuit, some pairs performed cursory evaluative measurements while others adopted a more extensive approach to evaluation. We used the Modeling Framework to gain insight into students' evaluation of the repaired circuit, as described below.

\subsection{Modeling}

All eight pairs engaged in model-based reasoning during either employment of the Split-Half Strategy to isolate Stage~2 as the source of faults (Pairs~D--H) and/or evaluation of the circuit's performance upon replacement of the op-amp in Stage~2 (Pairs~A--C and E--H). A summary of students' model-based reasoning during these episodes is given in Table~\ref{tab:modelingcodes}. While students engaged in a variety of Modeling phases (Sec.~\ref{sec:MF}) during one or both of these episodes, there are nevertheless differences in the nature of students' model-based reasoning in each case. For example, Model Construction and Prediction were \emph{more} common---and Proposal and Revision were \emph{less} common---during isolation of Stage~2 as a fault source than during replacement of the faulty op-amp.

\modelingtable

\subsubsection{Isolating Stage 2 as the fault source}

Five pairs (D--H) employed the Split-Half Strategy to isolate Stage~2 as the source of faults.  All five pairs made explicit statements in which they correctly identified Stage~1 as functional and/or isolated Stage~2 as the source of faults. For example, D2 concluded that ``the first one [Stage~1] is giving us a good voltage" and F1 said, ``The problem is in the second one [Stage~2]."

All five pairs engaged in Model Construction, Prediction, and Comparison (Table~\ref{tab:modelingcodes}). In all cases, Model Construction involved recognizing that the cascade amplifier consisted of two distinct stages, each characterized by a unique gain. Because Model Construction included recognizing that the overall gain of the circuit was the product of the gains of two stages, Model Construction was intertwined with Prediction. For example, after Students~D1 and D2 decided to measure the output of Stage~1, D2 said:
\QUOTE{And this [Stage~1] is noninverting and then [Stage~2 is] inverting. Oh yeah, that seems right. Because $\Vout$ equals negative this one [$R_4$] over that one [$R_3$]. Which is the second one [Stage~2]. So second one [Stage~2] should be negative ten [k\ohm] over one [k\ohm\ times]  $\Vin$. Which would be, which would be ten.}{D2}{34:59}{35:31}
D2 went on to clarify that in ``$\Vin$" in this instance meant ``$\Vout$ of the first one," \emph{i.e.}, the output of Stage~1. In this utterance, D2 first identified the existence of the two stages: ``this [Stage~1] is noninverting and then [Stage~2 is] inverting." D2 then identified and computed the gain of Stage~2: ``$\Vout$ equals negative this one [$R_4$] over that one [$R_3$] \ldots\ which would be ten." Thus, Model Construction and Prediction were occurring nearly simultaneously.

Similarly, Pair~E engaged in Model Construction, Prediction, and Comparison in short succession. After measuring the output of Stage~1, Student~E1 said:
\QUOTE{It looks like it's [Stage~1 is] doubling [the input signal]. So that seems right because the gain [of Stage~1] is one plus $R_2$ over $R_1$, which is one plus one. Two.}{E1}{19:36}{19:47}
E1 began by articulating the result of their measurement: the output of Stage~1 has twice the amplitude of the input to Stage~1 (``It looks like it's [Stage~1 is] doubling"). E1 then suggested that this measurement was in good agreement with the prediction (``that seems right") because the expected gain of Stage~1 was two. While Pair~E identified the existence of noninverting and inverting subsystems early on in the troubleshooting process, they did not identify algebraic expressions for the gains of either subsystem until this utterance. Thus the statement ``the gain is one plus $R_2$ over $R_1$" is an example of Model Construction, which was intertwined with both Comparison and Prediction.

Two pairs (D and F) engaged in Proposal. In addition to successfully eliminating Stage~1 as a potential source of faults, these pairs offered potential explanations for the discrepancy between expected and actual performance of the cascade amplifier: D2 suggested that the 1~k\ohm\ and 10~k\ohm\ resistors had accidentally been switched, and F1 suggested that the inputs to the op-amp in Stage~2 were incorrectly wired. While neither of these examples of Proposal yielded correct explanations for the observed discrepancies, both focused on the existence of potential faults in Stage~2.

\subsubsection{Replacing the faulty op-amp}\label{sec:evaluate}

Seven pairs (A--C and E--H) successfully repaired the circuit by replacing the op-amp in Stage~2. Here, we describe how these pairs engaged in all five aspects of the Modeling Framework listed in Table~\ref{tab:modelingcodes}.

Each of the seven pairs engaged in both Proposal and Revision: the students first proposed that the op-amp in Stage~2 was faulty and/or needed to be replaced, and then the students revised the circuit by replacing the faulty op-amp with a functional chip. For example, while brainstorming potential faults, Student~F1 said:
\QUOTE{Could the op-amps be faulty? Should we just replace them with new ones? 'Cause the second op amp---let's just replace it [the op-amp in Stage~2] with a new one. \ldots\ 'Cause the first one [Stage~1] is functioning fine.}{F1}{40:18}{40:38}
F1 initially proposed that faulty op-amps could be the cause of the malfunction, and suggested replacing them both with new chips. F1 then revised this proposal based on the results of earlier measurements, which revealed that the op-amp in Stage~1 was ``functioning fine." F1's revised proposal was to replace just the op-amp in Stage~2. Here, F1 used the results of the Split-Half Strategy to inform a proposed revision to the circuit.

Pair C used slightly different reasoning when proposing that the op-amp in Stage~2 was faulty. After performing diagnostic measurements of the voltage at several points in Stage~2, Student~C2 said:
\QUOTE{Pin 3 [the noninverting input terminal of the op-amp in Stage~2] is in fact zero. However, pin 2 [$V_{2-}$, the inverting input terminal of the op-amp in Stage~2] is not zero. And that's a problem. \ldots\ Certainly the second one [the op-amp in Stage~2] is not working because the Golden Rules are not being followed here.}{C2}{39:15}{39:34}
Here, C2 relied on their Domain Knowledge (\emph{i.e.}, their knowledge of the Golden Rule for op-amps in a closed loop) to identify the op-amp in Stage~2 as faulty.

Each of the seven pairs also engaged in Comparison: after replacing the faulty op-amp, the students performed evaluative measurements during which they compared the actual performance of the repaired circuit to their expectations. Four pairs (C and E--G) also engaged in Prediction by explicitly stating their expectations during the evaluation process.

The depth of evaluation varied among pairs of students. For example, after replacing both op-amps, Pair~B performed evaluative measurements of the amplitude and phase of the output of Stage~2. Upon performing these measurements, Student~B1 said:
\QUOTE{That was it. Nice. And now the output is 10 volts, maximum. And it's [the output signal is] inverted, which is good.}{B1}{46:35}{46:46}
B1 concluded that that the op-amps were indeed a fault source (``That was it."). B1 further articulated that the measured amplitude (``10 volts") and phase (``inverted") of the output signal were ``good," indicating satisfactory agreement with the expected amplitude and phase.

In addition to verifying the amplitude and phase of their output signal, Pair~G also checked to see that the new chip was satisfying the Golden Rule for op-amps in a closed loop:
\QUOTE{We should check the negative terminal [$V_{2-}$] here and it should be zero volts. \ldots\
[It is] point nine millivolts. So, that's basically zero. Okay. So, everything's behaving as it should now.}{G1}{32:22}{33:02}
Before measuring $V_{2-}$ of the new op-amp in Stage~2, G1 predicted that the voltage should be zero volts. G1 determined that the measured value, 0.9~mV, was ``basically zero" and that the circuit had been repaired (``everything's behaving as it should now."), indicating that the expected and actual performance of the circuit were in agreement.

Whereas Pair~G performed a more rigorous evaluation than Pair~B, Pair~A performed a more cursory evaluation. Like all other groups, Students~A1 and A2 measured the output of Stage~2 immediately after enacting their Revision. Because the amplitude of their ac input signal was larger than 650~mV, the output of the repaired circuit was a clipped ac signal, as shown in Plot~(d) of Fig.~\ref{fig:design}. After observing a clipped output signal, A2 remarked:
\QUOTE{That's just 'cause it's [the output signal is] railing. \ldots\ That's fine. Now, is the signal inverted? Yes, it is. Awesome. It works.}{A2}{21:18}{21:30}
A2 observed that the signal was being clipped (``it's railing"), but did not reduce the amplitude of the input signal in order to produce a sinusoidal output signal. This prevented Pair~A from determining the actual gain of the repaired circuit, which in turn prevented a quantitative comparison of the predicted and measured gains. Instead, A2 was satisfied (``That's fine.") with basic qualitative features of the output signal: it was an ac signal that was inverted with respect to the input. A2 concluded that the circuit has been repaired (``It works.").

Four pairs (A, C, E, and G) engaged in Model Construction during the evaluation of the repaired circuit. In all four cases, students articulated limitations of Eq.~(\ref{eq:transfer}). For example, after determining that their circuit had been repaired, Pair~E briefly discussed whether the observed behavior of the circuit made sense given the voltage limitations of the op-amps in the circuit:
\QUOTE{Let's see, how do we get 20 volts [for the output signal] when we only have 15 volts on the inputs [op-amp supply voltage]?}{E1}{28:22}{28:35}
E1 wondered how it was possible for the repaired circuit to provide an output of 20~V given that the positive and negative power rails of the op-amps were only $\pm15$~V. E2 noted that the output signal was 20~V from peak to peak, so the output was ``only going from minus 10 to plus 10." Thus, E1 and E2 engaged in Model Construction through articulation of model limitations.


\section{Discussion}\label{sec:discussion}

Our results demonstrate that each pair of students in our study engaged in all four cognitive troubleshooting tasks and used model-based reasoning during key strategic and/or evaluative phases of the troubleshooting activity.

Not only did all pairs engage in all four cognitive troubleshooting tasks, but the students were engaged in one or more tasks during almost every two-minute time interval throughout the activity. Moreover, the emergent patterns of engagement give rise to a sensible troubleshooting narrative with a beginning, middle, and end. In the beginning of the activity, students got their bearings on the problem by discerning the intended function of the circuit, performing visual inspections of the configuration, checking component values, replacing the faulty resistor, and making plans for how to test the circuit. Halfway through the activity, students began proposing potential faults, performing diagnostic measurements, and---in some cases---isolating the source of faults to the second stage of the circuit. Finally, at the end of the activity, almost all pairs identified the faulty op-amp, replaced it with a new chip, and favorably evaluated the performance of the repaired circuit.

While the cognitive tasks give us a coarse picture of students' approach to troubleshooting a malfunctioning circuit, the Modeling Framework allows us to look at two types of episodes in finer detail. All five pairs who employed the Split-Half Strategy engaged in model construction, prediction, and comparison. The Modeling Framework allows us to tell a sub-narrative about the Split-Half Strategy: the students first constructed a model of the circuit which consisted of two subsystems, each with its own gain; they were then able to form expectations about the outputs of the first and second stages; and, finally, by comparing their predictions to their measurements, the students successfully identified Stage~1 as functional and isolated Stage~2 as the source of faults.

Similarly, the Modeling Framework gives rise to a sub-narrative about the students' approach to evaluating the repaired circuit. In this case, all seven pairs who repaired the circuit engaged in proposal, revision, and comparison. The students first proposed a revision to the physical system apparatus, namely, replacing one or both of the op-amps with new chips; they then enacted this proposed revision; and, finally, the students compared the performance of the revised circuit to their expectations, thus ensuring that the circuit had indeed been repaired. In order to more thoroughly understand the performance of the repaired circuit, four pairs engaged in model construction by articulating constraints on the amplitude of the output voltage and checking to see that those constraints were being satisfied. For these students, model construction facilitated evaluation of their repairs.

The narratives above demonstrate the nonlinear, recursive nature of modeling. During the testing phase of the troubleshooting process, some students engaged in a modeling cycle to isolate the second stage as the source of faults. Later, another modeling cycle was needed to repair and evaluate the circuit. Furthermore, each modeling cycle had its own particular signature: pairs predominantly engaged in construction, prediction, and comparison in the former case compared to proposal, revision, and comparison in the latter. Thus, study participants engaged in multiple, distinct iterations of model-based reasoning while navigating the cognitive tasks required to troubleshoot a malfunctioning electrical circuit.

Because our participant pool was both small and homogenous (\emph{e.g.}, most participants were white men and both CU and UM are selective research-intensive institutions), our findings do not represent a comprehensive picture of students' approaches to troubleshooting electric circuits, nor do they necessarily speak to common or typical student responses to a troubleshooting activity. Rather, our findings show that the process of troubleshooting can engage students in the core scientific practice of modeling.


\section{Summary}\label{sec:summary}

We designed a think-aloud activity in which pairs of students attempted to repair a malfunctioning electrical circuit. The circuit was designed such that several troubleshooting strategies could be employed. Audiovisual data were collected for eight pairs of students from two different institutions. We used both a cognitive task analysis of troubleshooting and the Modeling Framework as \emph{a priori} schemes to analyze the data. Two types of episodes were chosen for in-depth analysis using the Modeling Framework: (1) isolation of one subsystem of the circuit as the source of faults, and (2) repair and evaluation of the circuit.

We found that all eight pairs engaged in all four cognitive troubleshooting tasks. Furthermore, in each of the two episodes chosen for in-depth analysis, we found a good mapping between students' actions and the Modeling Framework. Thus, we have shown that model-based reasoning facilitates the cognitive tasks required for troubleshooting. We have also demonstrated that the process of troubleshooting can engage students in the core scientific practice of modeling.

In ongoing work~\cite{VanDeBogart2015}, we are analyzing the data described here using a framework for socially mediated metacognition. Ultimately, we aim to use our understanding of the cognitive, metacognitive, and modeling-oriented aspects of troubleshooting to inform the development of activities to develop and assess students' troubleshooting abilities.

\acknowledgments The authors acknowledge the PER group at CU Boulder for useful input on study design and interpretation of results. This material is based upon work supported by the National Science Foundation under Grant Nos. DUE-1323101, DUE-1323426, DUE-1245313, and DUE-0962805.

\bibliography{ts_database}

\begin{thebibliography}{49}%
\makeatletter
\providecommand \@ifxundefined [1]{%
 \@ifx{#1\undefined}
}%
\providecommand \@ifnum [1]{%
 \ifnum #1\expandafter \@firstoftwo
 \else \expandafter \@secondoftwo
 \fi
}%
\providecommand \@ifx [1]{%
 \ifx #1\expandafter \@firstoftwo
 \else \expandafter \@secondoftwo
 \fi
}%
\providecommand \natexlab [1]{#1}%
\providecommand \enquote  [1]{``#1''}%
\providecommand \bibnamefont  [1]{#1}%
\providecommand \bibfnamefont [1]{#1}%
\providecommand \citenamefont [1]{#1}%
\providecommand \href@noop [0]{\@secondoftwo}%
\providecommand \href [0]{\begingroup \@sanitize@url \@href}%
\providecommand \@href[1]{\@@startlink{#1}\@@href}%
\providecommand \@@href[1]{\endgroup#1\@@endlink}%
\providecommand \@sanitize@url [0]{\catcode `\\12\catcode `\$12\catcode
  `\&12\catcode `\#12\catcode `\^12\catcode `\_12\catcode `\%12\relax}%
\providecommand \@@startlink[1]{}%
\providecommand \@@endlink[0]{}%
\providecommand \url  [0]{\begingroup\@sanitize@url \@url }%
\providecommand \@url [1]{\endgroup\@href {#1}{\urlprefix }}%
\providecommand \urlprefix  [0]{URL }%
\providecommand \Eprint [0]{\href }%
\providecommand \doibase [0]{http://dx.doi.org/}%
\providecommand \selectlanguage [0]{\@gobble}%
\providecommand \bibinfo  [0]{\@secondoftwo}%
\providecommand \bibfield  [0]{\@secondoftwo}%
\providecommand \translation [1]{[#1]}%
\providecommand \BibitemOpen [0]{}%
\providecommand \bibitemStop [0]{}%
\providecommand \bibitemNoStop [0]{.\EOS\space}%
\providecommand \EOS [0]{\spacefactor3000\relax}%
\providecommand \BibitemShut  [1]{\csname bibitem#1\endcsname}%
\let\auto@bib@innerbib\@empty
\bibitem [{\citenamefont {{National Research Council}}(2012)}]{DBER2012}%
  \BibitemOpen
  \bibfield  {author} {\bibinfo {author} {\bibnamefont {{National Research
  Council}}},\ }\href {http://www.nap.edu/catalog.php?record_id=13362} {\emph
  {\bibinfo {title} {{Discipline-Based Education Research: Understanding and
  Improving Learning in Undergraduate Science and Engineering}}}},\ edited by\
  \bibinfo {editor} {\bibfnamefont {Susan~R.}\ \bibnamefont {Singer}}, \bibinfo
  {editor} {\bibfnamefont {Natalie~R.}\ \bibnamefont {Nielsen}}, \ and\
  \bibinfo {editor} {\bibfnamefont {Heidi~A.}\ \bibnamefont {Schweingruber}}\
  (\bibinfo  {publisher} {National Academies Press},\ \bibinfo {year}
  {2012})\BibitemShut {NoStop}%
\bibitem [{\citenamefont {{PresidentÕs Council of Advisors on Science and
  Technology}}(2012)}]{PCAST2012}%
  \BibitemOpen
  \bibfield  {author} {\bibinfo {author} {\bibnamefont {{PresidentÕs Council of
  Advisors on Science and Technology}}},\ }\href@noop {} {\emph {\bibinfo
  {title} {{Engage to excel: Producing one million additional college graduates
  with degrees in science, technology, engineering, and mathematics}}}}\
  (\bibinfo  {publisher} {Executive Office of the President},\ \bibinfo {year}
  {2012})\BibitemShut {NoStop}%
\bibitem [{\citenamefont {{AAPT Committee on Laboratories}}(2015)}]{AAPT2015}%
  \BibitemOpen
  \bibfield  {author} {\bibinfo {author} {\bibnamefont {{AAPT Committee on
  Laboratories}}},\ }\href@noop {} {\emph {\bibinfo {title} {{AAPT
  Recommendations for the Undergraduate Physics Laboratory Curriculum}}}}\
  (\bibinfo  {publisher} {American Association of Physics Teachers},\ \bibinfo
  {year} {2015})\BibitemShut {NoStop}%
\bibitem [{\citenamefont {Wieman}(2015)}]{Wieman2015}%
  \BibitemOpen
  \bibfield  {author} {\bibinfo {author} {\bibfnamefont {Carl}\ \bibnamefont
  {Wieman}},\ }\bibfield  {title} {\enquote {\bibinfo {title} {Comparative
  cognitive task analyses of experimental science and instructional laboratory
  courses},}\ }\href {\doibase 10.1119/1.4928349} {\bibfield  {journal}
  {\bibinfo  {journal} {The Physics Teacher}\ }\textbf {\bibinfo {volume}
  {53}},\ \bibinfo {pages} {349--351} (\bibinfo {year} {2015})}\BibitemShut
  {NoStop}%
\bibitem [{\citenamefont {Zwickl}\ \emph {et~al.}(2015)\citenamefont {Zwickl},
  \citenamefont {Hu}, \citenamefont {Finkelstein},\ and\ \citenamefont
  {Lewandowski}}]{Zwickl2015}%
  \BibitemOpen
  \bibfield  {author} {\bibinfo {author} {\bibfnamefont {Benjamin~M.}\
  \bibnamefont {Zwickl}}, \bibinfo {author} {\bibfnamefont {Dehui}\
  \bibnamefont {Hu}}, \bibinfo {author} {\bibfnamefont {Noah}\ \bibnamefont
  {Finkelstein}}, \ and\ \bibinfo {author} {\bibfnamefont {H.~J.}\ \bibnamefont
  {Lewandowski}},\ }\bibfield  {title} {\enquote {\bibinfo {title} {Model-based
  reasoning in the physics laboratory: Framework and initial results},}\ }\href
  {\doibase 10.1103/PhysRevSTPER.11.020113} {\bibfield  {journal} {\bibinfo
  {journal} {Phys. Rev. ST Phys. Educ. Res.}\ }\textbf {\bibinfo {volume}
  {11}},\ \bibinfo {pages} {020113} (\bibinfo {year} {2015})}\BibitemShut
  {NoStop}%
\bibitem [{\citenamefont {{NGSS Lead States}}(2013)}]{NGSS2013}%
  \BibitemOpen
  \bibfield  {author} {\bibinfo {author} {\bibnamefont {{NGSS Lead States}}},\
  }\href@noop {} {\enquote {\bibinfo {title} {{Next Generation Science
  Standards: For States, By States}},}\ } (\bibinfo {year} {2013})\BibitemShut
  {NoStop}%
\bibitem [{\citenamefont {on~Undergraduate Physics Education~Research}\ \emph
  {et~al.}(2013)\citenamefont {on~Undergraduate Physics Education~Research},
  \citenamefont {on~Physics}, \citenamefont {on~Engineering},\ and\
  \citenamefont {Council}}]{NRC2013}%
  \BibitemOpen
  \bibfield  {author} {\bibinfo {author} {\bibfnamefont {Committee}\
  \bibnamefont {on~Undergraduate Physics Education~Research}}, \bibinfo
  {author} {\bibfnamefont {Implementation;~Board}\ \bibnamefont {on~Physics}},
  \bibinfo {author} {\bibfnamefont {Astronomy;~Division}\ \bibnamefont
  {on~Engineering}}, \ and\ \bibinfo {author} {\bibfnamefont {Physical
  Sciences; National~Research}\ \bibnamefont {Council}},\ }\href
  {http://www.nap.edu/openbook.php?record_id=18312} {\emph {\bibinfo {title}
  {Adapting to a Changing World--Challenges and Opportunities in Undergraduate
  Physics Education}}}\ (\bibinfo  {publisher} {The National Academies Press},\
  \bibinfo {year} {2013})\BibitemShut {NoStop}%
\bibitem [{\citenamefont {Etkina}\ and\ \citenamefont
  {Heuvelen}(2007)}]{Etkina2007}%
  \BibitemOpen
  \bibfield  {author} {\bibinfo {author} {\bibfnamefont {Eugenia}\ \bibnamefont
  {Etkina}}\ and\ \bibinfo {author} {\bibfnamefont {Alan~Van}\ \bibnamefont
  {Heuvelen}},\ }\bibfield  {title} {\enquote {\bibinfo {title} {Investigative
  science learning environment - a science process approach to learning
  physics},}\ }in\ \href@noop {} {\emph {\bibinfo {booktitle} {Research-Based
  Reform of University Physics}}},\ Vol.~\bibinfo {volume} {1}\ (\bibinfo
  {publisher} {American Association of Physics Teachers},\ \bibinfo {address}
  {College Park, MD},\ \bibinfo {year} {2007})\BibitemShut {NoStop}%
\bibitem [{\citenamefont {Brewe}(2008)}]{Brewe2008}%
  \BibitemOpen
  \bibfield  {author} {\bibinfo {author} {\bibfnamefont {Eric}\ \bibnamefont
  {Brewe}},\ }\bibfield  {title} {\enquote {\bibinfo {title} {Modeling theory
  applied: Modeling instruction in introductory physics},}\ }\href {\doibase
  http://dx.doi.org/10.1119/1.2983148} {\bibfield  {journal} {\bibinfo
  {journal} {American Journal of Physics}\ }\textbf {\bibinfo {volume} {76}},\
  \bibinfo {pages} {1155--1160} (\bibinfo {year} {2008})}\BibitemShut {NoStop}%
\bibitem [{\citenamefont {Zwickl}\ \emph {et~al.}(2012)\citenamefont {Zwickl},
  \citenamefont {Finkelstein},\ and\ \citenamefont {Lewandowski}}]{Zwickl2012}%
  \BibitemOpen
  \bibfield  {author} {\bibinfo {author} {\bibfnamefont {Benjamin~M.}\
  \bibnamefont {Zwickl}}, \bibinfo {author} {\bibfnamefont {Noah}\ \bibnamefont
  {Finkelstein}}, \ and\ \bibinfo {author} {\bibfnamefont {Heather~J.}\
  \bibnamefont {Lewandowski}},\ }\bibfield  {title} {\enquote {\bibinfo {title}
  {{Transforming the advanced lab: Part I - Learning goals}},}\ }\href
  {\doibase http://dx.doi.org/10.1063/1.3680077} {\bibfield  {journal}
  {\bibinfo  {journal} {AIP Conference Proceedings}\ }\textbf {\bibinfo
  {volume} {1413}},\ \bibinfo {pages} {391--394} (\bibinfo {year}
  {2012})}\BibitemShut {NoStop}%
\bibitem [{\citenamefont {Zwickl}\ \emph {et~al.}(2013)\citenamefont {Zwickl},
  \citenamefont {Finkelstein},\ and\ \citenamefont {Lewandowski}}]{Zwickl2013}%
  \BibitemOpen
  \bibfield  {author} {\bibinfo {author} {\bibfnamefont {Benjamin~M.}\
  \bibnamefont {Zwickl}}, \bibinfo {author} {\bibfnamefont {Noah}\ \bibnamefont
  {Finkelstein}}, \ and\ \bibinfo {author} {\bibfnamefont {Heather~J.}\
  \bibnamefont {Lewandowski}},\ }\bibfield  {title} {\enquote {\bibinfo {title}
  {The process of transforming an advanced lab course: Goals, curriculum, and
  assessments},}\ }\href {\doibase http://dx.doi.org/10.1119/1.4768890}
  {\bibfield  {journal} {\bibinfo  {journal} {American Journal of Physics}\
  }\textbf {\bibinfo {volume} {81}},\ \bibinfo {pages} {63--70} (\bibinfo
  {year} {2013})}\BibitemShut {NoStop}%
\bibitem [{\citenamefont {Lewandowski}\ and\ \citenamefont
  {Finkelstein}(2015)}]{Lewandowski2015}%
  \BibitemOpen
  \bibfield  {author} {\bibinfo {author} {\bibfnamefont {H.~J.}\ \bibnamefont
  {Lewandowski}}\ and\ \bibinfo {author} {\bibfnamefont {Noah}\ \bibnamefont
  {Finkelstein}},\ }\bibfield  {title} {\enquote {\bibinfo {title} {Redesigning
  a junior-level electronics course to support engagement in scientific
  practices},}\ }in\ \href {\doibase 10.1119/perc.2015.pr.043} {\emph {\bibinfo
  {booktitle} {Physics Education Research Conference 2015}}},\ \bibinfo {series
  and number} {PER Conference}\ (\bibinfo {address} {College Park, MD},\
  \bibinfo {year} {2015})\ pp.\ \bibinfo {pages} {191--194}\BibitemShut
  {NoStop}%
\bibitem [{\citenamefont {Perez}(1991)}]{Perez1991}%
  \BibitemOpen
  \bibfield  {author} {\bibinfo {author} {\bibfnamefont {Ray~S.}\ \bibnamefont
  {Perez}},\ }\bibfield  {title} {\enquote {\bibinfo {title} {A view from
  troubleshooting},}\ }in\ \href@noop {} {\emph {\bibinfo {booktitle} {Toward a
  Unified Theory of Problem Solving: Views From the Content Domains}}},\
  \bibinfo {editor} {edited by\ \bibinfo {editor} {\bibfnamefont {Mike~U.}\
  \bibnamefont {Smith}}}\ (\bibinfo  {publisher} {Taylor \& Francis Group},\
  \bibinfo {address} {New York},\ \bibinfo {year} {1991})\ pp.\ \bibinfo
  {pages} {115--153}\BibitemShut {NoStop}%
\bibitem [{\citenamefont {Schaafstal}\ \emph {et~al.}(2000)\citenamefont
  {Schaafstal}, \citenamefont {Schraagen},\ and\ \citenamefont
  {Van~Berl}}]{Schaafstal2000}%
  \BibitemOpen
  \bibfield  {author} {\bibinfo {author} {\bibfnamefont {Alma}\ \bibnamefont
  {Schaafstal}}, \bibinfo {author} {\bibfnamefont {Jan~Maarten}\ \bibnamefont
  {Schraagen}}, \ and\ \bibinfo {author} {\bibfnamefont {Marcel}\ \bibnamefont
  {Van~Berl}},\ }\bibfield  {title} {\enquote {\bibinfo {title} {Cognitive task
  analysis and innovation of training: The case of structured
  troubleshooting},}\ }\href {\doibase 10.1518/001872000779656570} {\bibfield
  {journal} {\bibinfo  {journal} {Human Factors: The Journal of the Human
  Factors and Ergonomics Society}\ }\textbf {\bibinfo {volume} {42}},\ \bibinfo
  {pages} {75--86} (\bibinfo {year} {2000})}\BibitemShut {NoStop}%
\bibitem [{\citenamefont {Jonassen}\ and\ \citenamefont
  {Hung}(2006)}]{Jonassen2006}%
  \BibitemOpen
  \bibfield  {author} {\bibinfo {author} {\bibfnamefont {David~H.}\
  \bibnamefont {Jonassen}}\ and\ \bibinfo {author} {\bibfnamefont {Woei}\
  \bibnamefont {Hung}},\ }\bibfield  {title} {\enquote {\bibinfo {title}
  {Learning to troubleshoot: A new theory-based design architecture},}\ }\href
  {\doibase 10.1007/s10648-006-9001-8} {\bibfield  {journal} {\bibinfo
  {journal} {Educational Psychology Review}\ }\textbf {\bibinfo {volume}
  {18}},\ \bibinfo {pages} {77--114} (\bibinfo {year} {2006})}\BibitemShut
  {NoStop}%
\bibitem [{\citenamefont {MacPherson}(1998)}]{MacPherson1998}%
  \BibitemOpen
  \bibfield  {author} {\bibinfo {author} {\bibfnamefont {Randall~T.}\
  \bibnamefont {MacPherson}},\ }\bibfield  {title} {\enquote {\bibinfo {title}
  {Factors affecting technological trouble shooting skills},}\ }\href
  {http://scholar.lib.vt.edu/ejournals/JITE/v35n4/macpherson.html} {\bibfield
  {journal} {\bibinfo  {journal} {Journal of Industrial Teacher Education}\
  }\textbf {\bibinfo {volume} {35}},\ \bibinfo {pages} {5--28} (\bibinfo {year}
  {1998})}\BibitemShut {NoStop}%
\bibitem [{\citenamefont {Johnson}(1988)}]{Johnson1988}%
  \BibitemOpen
  \bibfield  {author} {\bibinfo {author} {\bibfnamefont {Scott~D.}\
  \bibnamefont {Johnson}},\ }\bibfield  {title} {\enquote {\bibinfo {title}
  {Cognitive analysis of expert and novice troubleshooting performance},}\
  }\href {\doibase 10.1111/j.1937-8327.1988.tb00021.x} {\bibfield  {journal}
  {\bibinfo  {journal} {Performance Improvement Quarterly}\ }\textbf {\bibinfo
  {volume} {1}},\ \bibinfo {pages} {38--54} (\bibinfo {year}
  {1988})}\BibitemShut {NoStop}%
\bibitem [{\citenamefont {Van~Gog}\ \emph
  {et~al.}(2005{\natexlab{a}})\citenamefont {Van~Gog}, \citenamefont {Paas},
  \citenamefont {Van~Merri\"enboer},\ and\ \citenamefont
  {Witte}}]{vanGog2005a}%
  \BibitemOpen
  \bibfield  {author} {\bibinfo {author} {\bibfnamefont {Tamara}\ \bibnamefont
  {Van~Gog}}, \bibinfo {author} {\bibfnamefont {Fred}\ \bibnamefont {Paas}},
  \bibinfo {author} {\bibfnamefont {Jeroen J.~G.}\ \bibnamefont
  {Van~Merri\"enboer}}, \ and\ \bibinfo {author} {\bibfnamefont {Puk}\
  \bibnamefont {Witte}},\ }\bibfield  {title} {\enquote {\bibinfo {title}
  {Uncovering the problem-solving process: Cued retrospective reporting versus
  concurrent and retrospective reporting},}\ }\href {\doibase
  10.1037/1076-898X.11.4.237} {\bibfield  {journal} {\bibinfo  {journal}
  {Journal of Experimental Psychology: Applied}\ }\textbf {\bibinfo {volume}
  {11}},\ \bibinfo {pages} {237--244} (\bibinfo {year}
  {2005}{\natexlab{a}})}\BibitemShut {NoStop}%
\bibitem [{\citenamefont {Van~Gog}\ \emph
  {et~al.}(2005{\natexlab{b}})\citenamefont {Van~Gog}, \citenamefont {Paas},\
  and\ \citenamefont {Van~Merri\"enboer}}]{vanGog2005b}%
  \BibitemOpen
  \bibfield  {author} {\bibinfo {author} {\bibfnamefont {Tamara}\ \bibnamefont
  {Van~Gog}}, \bibinfo {author} {\bibfnamefont {Fred}\ \bibnamefont {Paas}}, \
  and\ \bibinfo {author} {\bibfnamefont {Jeroen J.~G.}\ \bibnamefont
  {Van~Merri\"enboer}},\ }\bibfield  {title} {\enquote {\bibinfo {title}
  {Uncovering expertise-related differences in troubleshooting performance:
  combining eye movement and concurrent verbal protocol data},}\ }\href
  {\doibase 10.1002/acp.1112} {\bibfield  {journal} {\bibinfo  {journal}
  {Applied Cognitive Psychology}\ }\textbf {\bibinfo {volume} {19}},\ \bibinfo
  {pages} {205--221} (\bibinfo {year} {2005}{\natexlab{b}})}\BibitemShut
  {NoStop}%
\bibitem [{\citenamefont {Gog}\ \emph {et~al.}(2006)\citenamefont {Gog},
  \citenamefont {Paas},\ and\ \citenamefont {Merri\"enboer}}]{vanGog2006}%
  \BibitemOpen
  \bibfield  {author} {\bibinfo {author} {\bibfnamefont {Tamara~Van}\
  \bibnamefont {Gog}}, \bibinfo {author} {\bibfnamefont {Fred}\ \bibnamefont
  {Paas}}, \ and\ \bibinfo {author} {\bibfnamefont {Jeroen J. G.~Van}\
  \bibnamefont {Merri\"enboer}},\ }\bibfield  {title} {\enquote {\bibinfo
  {title} {Effects of process-oriented worked examples on troubleshooting
  transfer performance},}\ }\href {\doibase
  http://dx.doi.org/10.1016/j.learninstruc.2006.02.003} {\bibfield  {journal}
  {\bibinfo  {journal} {Learning and Instruction}\ }\textbf {\bibinfo {volume}
  {16}},\ \bibinfo {pages} {154--164} (\bibinfo {year} {2006})}\BibitemShut
  {NoStop}%
\bibitem [{\citenamefont {Gog}\ \emph {et~al.}(2008)\citenamefont {Gog},
  \citenamefont {Paas},\ and\ \citenamefont {Merri\"enboer}}]{vanGog2008}%
  \BibitemOpen
  \bibfield  {author} {\bibinfo {author} {\bibfnamefont {Tamara~Van}\
  \bibnamefont {Gog}}, \bibinfo {author} {\bibfnamefont {Fred}\ \bibnamefont
  {Paas}}, \ and\ \bibinfo {author} {\bibfnamefont {Jeroen J. G.~Van}\
  \bibnamefont {Merri\"enboer}},\ }\bibfield  {title} {\enquote {\bibinfo
  {title} {Effects of studying sequences of process-oriented and
  product-oriented worked examples on troubleshooting transfer efficiency},}\
  }\href {\doibase http://dx.doi.org/10.1016/j.learninstruc.2007.03.003}
  {\bibfield  {journal} {\bibinfo  {journal} {Learning and Instruction}\
  }\textbf {\bibinfo {volume} {18}},\ \bibinfo {pages} {211--222} (\bibinfo
  {year} {2008})}\BibitemShut {NoStop}%
\bibitem [{\citenamefont {Kester}\ \emph {et~al.}(2004)\citenamefont {Kester},
  \citenamefont {Kirschner},\ and\ \citenamefont {Merri\"enboer}}]{Kester2004}%
  \BibitemOpen
  \bibfield  {author} {\bibinfo {author} {\bibfnamefont {Liesbeth}\
  \bibnamefont {Kester}}, \bibinfo {author} {\bibfnamefont {Paul~A.}\
  \bibnamefont {Kirschner}}, \ and\ \bibinfo {author} {\bibfnamefont {Jeroen J.
  G.~Van}\ \bibnamefont {Merri\"enboer}},\ }\bibfield  {title} {\enquote
  {\bibinfo {title} {Information presentation and troubleshooting in electrical
  circuits},}\ }\href {\doibase 10.1080/69032000072809} {\bibfield  {journal}
  {\bibinfo  {journal} {International Journal of Science Education}\ }\textbf
  {\bibinfo {volume} {26}},\ \bibinfo {pages} {239--256} (\bibinfo {year}
  {2004})}\BibitemShut {NoStop}%
\bibitem [{\citenamefont {Kester}\ \emph {et~al.}(2006)\citenamefont {Kester},
  \citenamefont {Kirschner},\ and\ \citenamefont {Merri\"enboer}}]{Kester2006}%
  \BibitemOpen
  \bibfield  {author} {\bibinfo {author} {\bibfnamefont {Liesbeth}\
  \bibnamefont {Kester}}, \bibinfo {author} {\bibfnamefont {Paul~A.}\
  \bibnamefont {Kirschner}}, \ and\ \bibinfo {author} {\bibfnamefont {Jeroen J.
  G.~Van}\ \bibnamefont {Merri\"enboer}},\ }\bibfield  {title} {\enquote
  {\bibinfo {title} {Just-in-time information presentation: Improving learning
  a troubleshooting skill},}\ }\href {\doibase
  http://dx.doi.org/10.1016/j.cedpsych.2005.04.002} {\bibfield  {journal}
  {\bibinfo  {journal} {Contemporary Educational Psychology}\ }\textbf
  {\bibinfo {volume} {31}},\ \bibinfo {pages} {167--185} (\bibinfo {year}
  {2006})}\BibitemShut {NoStop}%
\bibitem [{\citenamefont {Johnson}\ and\ \citenamefont
  {Chung}(1999)}]{Johnson1999}%
  \BibitemOpen
  \bibfield  {author} {\bibinfo {author} {\bibfnamefont {Scott~D.}\
  \bibnamefont {Johnson}}\ and\ \bibinfo {author} {\bibfnamefont {Shih-Ping}\
  \bibnamefont {Chung}},\ }\bibfield  {title} {\enquote {\bibinfo {title} {The
  effect of thinking aloud pair problem solving (tapps) on the troubleshooting
  ability of aviation technician students},}\ }\href
  {http://scholar.lib.vt.edu/ejournals/JITE/v37n1/john.html} {\bibfield
  {journal} {\bibinfo  {journal} {Journal of Industrial Teacher Education}\
  }\textbf {\bibinfo {volume} {37}} (\bibinfo {year} {1999})}\BibitemShut
  {NoStop}%
\bibitem [{\citenamefont {Ross}\ and\ \citenamefont {Orr}(2007)}]{Ross2007}%
  \BibitemOpen
  \bibfield  {author} {\bibinfo {author} {\bibfnamefont {Craig}\ \bibnamefont
  {Ross}}\ and\ \bibinfo {author} {\bibfnamefont {R.~Robert}\ \bibnamefont
  {Orr}},\ }\bibfield  {title} {\enquote {\bibinfo {title} {Teaching structured
  troubleshooting: integrating a standard methodology into an information
  technology program},}\ }\href {\doibase 10.1007/s11423-007-9047-4} {\bibfield
   {journal} {\bibinfo  {journal} {Educational Technology Research and
  Development}\ }\textbf {\bibinfo {volume} {57}},\ \bibinfo {pages} {251--265}
  (\bibinfo {year} {2007})}\BibitemShut {NoStop}%
\bibitem [{\citenamefont {McDermott}\ and\ \citenamefont
  {Shaffer}(1992)}]{McDermott1992}%
  \BibitemOpen
  \bibfield  {author} {\bibinfo {author} {\bibfnamefont {Lillian~C.}\
  \bibnamefont {McDermott}}\ and\ \bibinfo {author} {\bibfnamefont {Peter~S.}\
  \bibnamefont {Shaffer}},\ }\bibfield  {title} {\enquote {\bibinfo {title}
  {{Research as a guide for curriculum development: An example from
  introductory electricity. Part I: Investigation of student understanding}},}\
  }\href {\doibase http://dx.doi.org/10.1119/1.17003} {\bibfield  {journal}
  {\bibinfo  {journal} {American Journal of Physics}\ }\textbf {\bibinfo
  {volume} {60}},\ \bibinfo {pages} {994--1003} (\bibinfo {year}
  {1992})}\BibitemShut {NoStop}%
\bibitem [{\citenamefont {McDermott}\ and\ \citenamefont
  {Shaffer}(1993)}]{McDermott1993}%
  \BibitemOpen
  \bibfield  {author} {\bibinfo {author} {\bibfnamefont {Lillian~C.}\
  \bibnamefont {McDermott}}\ and\ \bibinfo {author} {\bibfnamefont {Peter~S.}\
  \bibnamefont {Shaffer}},\ }\bibfield  {title} {\enquote {\bibinfo {title}
  {{Erratum: ÔÔResearch as a guide for curriculum development: An example from
  introductory electricity. Part I: Investigation of student understandingÕÕ
  [Am. J. Phys. 60, 994Ð1003 (1992)]}},}\ }\href {\doibase
  http://dx.doi.org/10.1119/1.17448} {\bibfield  {journal} {\bibinfo  {journal}
  {American Journal of Physics}\ }\textbf {\bibinfo {volume} {61}},\ \bibinfo
  {pages} {81--81} (\bibinfo {year} {1993})}\BibitemShut {NoStop}%
\bibitem [{\citenamefont {Engelhardt}\ and\ \citenamefont
  {Beichner}(2004)}]{Engelhardt2004}%
  \BibitemOpen
  \bibfield  {author} {\bibinfo {author} {\bibfnamefont {Paula~Vetter}\
  \bibnamefont {Engelhardt}}\ and\ \bibinfo {author} {\bibfnamefont
  {Robert~J.}\ \bibnamefont {Beichner}},\ }\bibfield  {title} {\enquote
  {\bibinfo {title} {StudentsÕ understanding of direct current resistive
  electrical circuits},}\ }\href {\doibase http://dx.doi.org/10.1119/1.1614813}
  {\bibfield  {journal} {\bibinfo  {journal} {American Journal of Physics}\
  }\textbf {\bibinfo {volume} {72}},\ \bibinfo {pages} {98--115} (\bibinfo
  {year} {2004})}\BibitemShut {NoStop}%
\bibitem [{\citenamefont {Coppens}\ \emph {et~al.}(2012)\citenamefont
  {Coppens}, \citenamefont {De~Cock},\ and\ \citenamefont
  {Kautz}}]{Coppens2012}%
  \BibitemOpen
  \bibfield  {author} {\bibinfo {author} {\bibfnamefont {P.}~\bibnamefont
  {Coppens}}, \bibinfo {author} {\bibfnamefont {M.}~\bibnamefont {De~Cock}}, \
  and\ \bibinfo {author} {\bibfnamefont {C.}~\bibnamefont {Kautz}},\ }\bibfield
   {title} {\enquote {\bibinfo {title} {Student understanding of filters in
  analog electronics lab courses},}\ }in\ \href@noop {} {\emph {\bibinfo
  {booktitle} {Proceedings of the 40th SEFI Annual Conference 2012}}}\
  (\bibinfo {address} {Thessaloniki, Greece},\ \bibinfo {year}
  {2012})\BibitemShut {NoStop}%
\bibitem [{\citenamefont {Stetzer}\ \emph {et~al.}(2013)\citenamefont
  {Stetzer}, \citenamefont {Van~Kampen}, \citenamefont {Shaffer},\ and\
  \citenamefont {McDermott}}]{Stetzer2013}%
  \BibitemOpen
  \bibfield  {author} {\bibinfo {author} {\bibfnamefont {MacKenzie~R.}\
  \bibnamefont {Stetzer}}, \bibinfo {author} {\bibfnamefont {Paul}\
  \bibnamefont {Van~Kampen}}, \bibinfo {author} {\bibfnamefont {Peter~S.}\
  \bibnamefont {Shaffer}}, \ and\ \bibinfo {author} {\bibfnamefont
  {Lillian~C.}\ \bibnamefont {McDermott}},\ }\bibfield  {title} {\enquote
  {\bibinfo {title} {New insights into student understanding of complete
  circuits and the conservation of current},}\ }\href {\doibase
  http://dx.doi.org/10.1119/1.4773293} {\bibfield  {journal} {\bibinfo
  {journal} {American Journal of Physics}\ }\textbf {\bibinfo {volume} {81}},\
  \bibinfo {pages} {134--143} (\bibinfo {year} {2013})}\BibitemShut {NoStop}%
\bibitem [{\citenamefont {Papanikolaou}\ \emph {et~al.}(2015)\citenamefont
  {Papanikolaou}, \citenamefont {Tombras}, \citenamefont {Van De~Bogart},\ and\
  \citenamefont {Stetzer}}]{Papanikolaou2015}%
  \BibitemOpen
  \bibfield  {author} {\bibinfo {author} {\bibfnamefont {Christos~P.}\
  \bibnamefont {Papanikolaou}}, \bibinfo {author} {\bibfnamefont {George~S.}\
  \bibnamefont {Tombras}}, \bibinfo {author} {\bibfnamefont {Kevin~L.}\
  \bibnamefont {Van De~Bogart}}, \ and\ \bibinfo {author} {\bibfnamefont
  {MacKenzie~R.}\ \bibnamefont {Stetzer}},\ }\bibfield  {title} {\enquote
  {\bibinfo {title} {Investigating student understanding of
  operational-amplifier circuits},}\ }\href {\doibase
  http://dx.doi.org/10.1119/1.4934600} {\bibfield  {journal} {\bibinfo
  {journal} {American Journal of Physics}\ }\textbf {\bibinfo {volume} {83}},\
  \bibinfo {pages} {1039--1050} (\bibinfo {year} {2015})}\BibitemShut {NoStop}%
\bibitem [{\citenamefont {Shaffer}\ and\ \citenamefont
  {McDermott}(1992)}]{Shaffer1992}%
  \BibitemOpen
  \bibfield  {author} {\bibinfo {author} {\bibfnamefont {Peter~S.}\
  \bibnamefont {Shaffer}}\ and\ \bibinfo {author} {\bibfnamefont {Lillian~C.}\
  \bibnamefont {McDermott}},\ }\bibfield  {title} {\enquote {\bibinfo {title}
  {{Research as a guide for curriculum development: An example from
  introductory electricity. Part II: Design of instructional strategies}},}\
  }\href {\doibase http://dx.doi.org/10.1119/1.16979} {\bibfield  {journal}
  {\bibinfo  {journal} {American Journal of Physics}\ }\textbf {\bibinfo
  {volume} {60}},\ \bibinfo {pages} {1003--1013} (\bibinfo {year}
  {1992})}\BibitemShut {NoStop}%
\bibitem [{\citenamefont {Getty}(2009)}]{Getty2009}%
  \BibitemOpen
  \bibfield  {author} {\bibinfo {author} {\bibfnamefont {J.C.}\ \bibnamefont
  {Getty}},\ }\bibfield  {title} {\enquote {\bibinfo {title} {Assessing inquiry
  learning in a circuits/electronics course},}\ }in\ \href@noop {} {\emph
  {\bibinfo {booktitle} {Frontiers in Education Conference, 2009. FIE '09. 39th
  IEEE}}}\ (\bibinfo {year} {2009})\ pp.\ \bibinfo {pages} {1--6}\BibitemShut
  {NoStop}%
\bibitem [{\citenamefont {Mazzolini}\ \emph {et~al.}(2011)\citenamefont
  {Mazzolini}, \citenamefont {Edwards}, \citenamefont {Rachinger},\ and\
  \citenamefont {Nopparatjamjomras}}]{Mazzolini2011}%
  \BibitemOpen
  \bibfield  {author} {\bibinfo {author} {\bibfnamefont {A.}~\bibnamefont
  {Mazzolini}}, \bibinfo {author} {\bibfnamefont {T.}~\bibnamefont {Edwards}},
  \bibinfo {author} {\bibfnamefont {W.}~\bibnamefont {Rachinger}}, \ and\
  \bibinfo {author} {\bibfnamefont {S.}~\bibnamefont {Nopparatjamjomras}},\
  }\bibfield  {title} {\enquote {\bibinfo {title} {The use of interactive
  lecture demonstrations to improve students' understanding of operational
  amplifiers in a tertiary introductory electronics course},}\ }\href@noop {}
  {\bibfield  {journal} {\bibinfo  {journal} {Latin-American Journal of Physics
  Education}\ }\textbf {\bibinfo {volume} {5}},\ \bibinfo {pages} {147--153}
  (\bibinfo {year} {2011})}\BibitemShut {NoStop}%
\bibitem [{\citenamefont {Dounas-Frazer}\ \emph {et~al.}(2015)\citenamefont
  {Dounas-Frazer}, \citenamefont {Bogart}, \citenamefont {Stetzer},\ and\
  \citenamefont {Lewandowski}}]{Dounas-Frazer2015}%
  \BibitemOpen
  \bibfield  {author} {\bibinfo {author} {\bibfnamefont {Dimitri}\ \bibnamefont
  {Dounas-Frazer}}, \bibinfo {author} {\bibfnamefont {Kevin Van~De}\
  \bibnamefont {Bogart}}, \bibinfo {author} {\bibfnamefont {MacKenzie~R.}\
  \bibnamefont {Stetzer}}, \ and\ \bibinfo {author} {\bibfnamefont {H.~J.}\
  \bibnamefont {Lewandowski}},\ }\bibfield  {title} {\enquote {\bibinfo {title}
  {The role of modeling in troubleshooting: An example from electronics},}\
  }in\ \href {\doibase 10.1119/perc.2015.pr.021} {\emph {\bibinfo {booktitle}
  {Physics Education Research Conference 2015}}},\ \bibinfo {series and number}
  {PER Conference}\ (\bibinfo {address} {College Park, MD},\ \bibinfo {year}
  {2015})\ pp.\ \bibinfo {pages} {103--106}\BibitemShut {NoStop}%
\bibitem [{\citenamefont {Bogart}\ \emph {et~al.}(2015)\citenamefont {Bogart},
  \citenamefont {Dounas-Frazer}, \citenamefont {Lewandowski},\ and\
  \citenamefont {Stetzer}}]{VanDeBogart2015}%
  \BibitemOpen
  \bibfield  {author} {\bibinfo {author} {\bibfnamefont {Kevin Van~De}\
  \bibnamefont {Bogart}}, \bibinfo {author} {\bibfnamefont {Dimitri}\
  \bibnamefont {Dounas-Frazer}}, \bibinfo {author} {\bibfnamefont {H.~J.}\
  \bibnamefont {Lewandowski}}, \ and\ \bibinfo {author} {\bibfnamefont
  {MacKenzie~R.}\ \bibnamefont {Stetzer}},\ }\bibfield  {title} {\enquote
  {\bibinfo {title} {The role of metacognition in troubleshooting: An example
  from electronics},}\ }in\ \href {\doibase 10.1119/perc.2015.pr.080} {\emph
  {\bibinfo {booktitle} {Physics Education Research Conference 2015}}},\
  \bibinfo {series and number} {PER Conference}\ (\bibinfo {address} {College
  Park, MD},\ \bibinfo {year} {2015})\ pp.\ \bibinfo {pages}
  {339--342}\BibitemShut {NoStop}%
\bibitem [{\citenamefont {Crandall}\ \emph {et~al.}(2006)\citenamefont
  {Crandall}, \citenamefont {Klein},\ and\ \citenamefont
  {Hoffman}}]{Crandall2006}%
  \BibitemOpen
  \bibfield  {author} {\bibinfo {author} {\bibfnamefont {Beth}\ \bibnamefont
  {Crandall}}, \bibinfo {author} {\bibfnamefont {Gary}\ \bibnamefont {Klein}},
  \ and\ \bibinfo {author} {\bibfnamefont {Robert~R.}\ \bibnamefont
  {Hoffman}},\ }\href@noop {} {\emph {\bibinfo {title} {Working minds: A
  practitioner's guide to {Cognitive Task Analysis}}}}\ (\bibinfo  {publisher}
  {MIT Press},\ \bibinfo {address} {Cambridge},\ \bibinfo {year}
  {2006})\BibitemShut {NoStop}%
\bibitem [{\citenamefont {Davis}(1983)}]{Davis1983}%
  \BibitemOpen
  \bibfield  {author} {\bibinfo {author} {\bibfnamefont {Randall}\ \bibnamefont
  {Davis}},\ }\bibfield  {title} {\enquote {\bibinfo {title} {Reasoning from
  first principles in electronic troubleshooting},}\ }\href {\doibase
  http://dx.doi.org/10.1016/S0020-7373(83)80063-7} {\bibfield  {journal}
  {\bibinfo  {journal} {International Journal of Man-Machine Studies}\ }\textbf
  {\bibinfo {volume} {19}},\ \bibinfo {pages} {403--423} (\bibinfo {year}
  {1983})}\BibitemShut {NoStop}%
\bibitem [{\citenamefont {Goos}\ \emph {et~al.}(2002)\citenamefont {Goos},
  \citenamefont {Galbraith},\ and\ \citenamefont {Renshaw}}]{Goos2002}%
  \BibitemOpen
  \bibfield  {author} {\bibinfo {author} {\bibfnamefont {Merrilyn}\
  \bibnamefont {Goos}}, \bibinfo {author} {\bibfnamefont {Peter}\ \bibnamefont
  {Galbraith}}, \ and\ \bibinfo {author} {\bibfnamefont {Peter}\ \bibnamefont
  {Renshaw}},\ }\bibfield  {title} {\enquote {\bibinfo {title} {Socially
  mediated metacognition: creating collaborative zones of proximal development
  in small group problem solving},}\ }\href {\doibase 10.1023/A:1016209010120}
  {\bibfield  {journal} {\bibinfo  {journal} {Educational Studies in
  Mathematics}\ }\textbf {\bibinfo {volume} {49}},\ \bibinfo {pages} {193--223}
  (\bibinfo {year} {2002})}\BibitemShut {NoStop}%
\bibitem [{\citenamefont {Bereiter}\ and\ \citenamefont
  {Miller}(1989)}]{Bereiter1989}%
  \BibitemOpen
  \bibfield  {author} {\bibinfo {author} {\bibfnamefont {S.R.}\ \bibnamefont
  {Bereiter}}\ and\ \bibinfo {author} {\bibfnamefont {S.M.}\ \bibnamefont
  {Miller}},\ }\bibfield  {title} {\enquote {\bibinfo {title} {A field-based
  study of troubleshooting in computer-controlled manufacturing systems},}\
  }\href {\doibase 10.1109/21.31027} {\bibfield  {journal} {\bibinfo  {journal}
  {Systems, Man and Cybernetics, IEEE Transactions on}\ }\textbf {\bibinfo
  {volume} {19}},\ \bibinfo {pages} {205--219} (\bibinfo {year}
  {1989})}\BibitemShut {NoStop}%
\bibitem [{\citenamefont {Konradt}(1995)}]{Konradt1995}%
  \BibitemOpen
  \bibfield  {author} {\bibinfo {author} {\bibfnamefont {Udo}\ \bibnamefont
  {Konradt}},\ }\bibfield  {title} {\enquote {\bibinfo {title} {Strategies of
  failure diagnosis in computer-controlled manufacturing systems: empirical
  analysis and implications for the design of adaptive decision support
  systems},}\ }\href {\doibase http://dx.doi.org/10.1006/ijhc.1995.1057}
  {\bibfield  {journal} {\bibinfo  {journal} {International Journal of
  Human-Computer Studies}\ }\textbf {\bibinfo {volume} {43}},\ \bibinfo {pages}
  {503 -- 521} (\bibinfo {year} {1995})}\BibitemShut {NoStop}%
\bibitem [{\citenamefont {Redish}(2002)}]{Redish2002}%
  \BibitemOpen
  \bibfield  {author} {\bibinfo {author} {\bibfnamefont {Edward~F.}\
  \bibnamefont {Redish}},\ }\href {http://www2.physics.umd.edu/~redish/Book}
  {\emph {\bibinfo {title} {Teaching Physics with the Physics Suite}}}\
  (\bibinfo  {publisher} {University of Maryland Physics Education Research
  Group},\ \bibinfo {address} {College Park},\ \bibinfo {year}
  {2002})\BibitemShut {NoStop}%
\bibitem [{\citenamefont {{Carnegie Foundation for the Advancement of
  Teaching}}(2011)}]{Carnegie2011}%
  \BibitemOpen
  \bibfield  {author} {\bibinfo {author} {\bibnamefont {{Carnegie Foundation
  for the Advancement of Teaching}}},\ }\href@noop {} {\emph {\bibinfo {title}
  {{The Carnegie Classification of Institutions of Higher Education, 2010
  edition}}}}\ (\bibinfo {address} {Menlo Park, CA},\ \bibinfo {year}
  {2011})\BibitemShut {NoStop}%
\bibitem [{\citenamefont {Lewandowski}\ \emph {et~al.}(2014)\citenamefont
  {Lewandowski}, \citenamefont {Finkelstein},\ and\ \citenamefont
  {Pollard}}]{Pollard2014}%
  \BibitemOpen
  \bibfield  {author} {\bibinfo {author} {\bibfnamefont {H.~J.}\ \bibnamefont
  {Lewandowski}}, \bibinfo {author} {\bibfnamefont {Noah}\ \bibnamefont
  {Finkelstein}}, \ and\ \bibinfo {author} {\bibfnamefont {Benjamin}\
  \bibnamefont {Pollard}},\ }\bibfield  {title} {\enquote {\bibinfo {title}
  {Studying expert practices to create learning goals for electronics labs},}\
  }in\ \href {\doibase 10.1119/perc.2014.pr.035} {\emph {\bibinfo {booktitle}
  {Physics Education Research Conference 2014}}},\ \bibinfo {series and number}
  {PER Conference}\ (\bibinfo {address} {Minneapolis, MN},\ \bibinfo {year}
  {2014})\ pp.\ \bibinfo {pages} {155--158}\BibitemShut {NoStop}%
\bibitem [{\citenamefont {{Ericsson, K. A., and Simon, H.
  A}}(1993)}]{Ericsson1993}%
  \BibitemOpen
  \bibfield  {author} {\bibinfo {author} {\bibnamefont {{Ericsson, K. A., and
  Simon, H. A}}},\ }\href@noop {} {\emph {\bibinfo {title} {{Protocol analysis:
  Verbal reports as data (Rev. ed.)}}}}\ (\bibinfo  {publisher} {MIT Press},\
  \bibinfo {address} {Cambridge, MA},\ \bibinfo {year} {1993})\BibitemShut
  {NoStop}%
\bibitem [{\citenamefont {Van~Someren}\ \emph {et~al.}(1994)\citenamefont
  {Van~Someren}, \citenamefont {Barnard},\ and\ \citenamefont
  {Sandberg}}]{vanSomeren1994}%
  \BibitemOpen
  \bibfield  {author} {\bibinfo {author} {\bibfnamefont {M.~W.}\ \bibnamefont
  {Van~Someren}}, \bibinfo {author} {\bibfnamefont {Y.~F.}\ \bibnamefont
  {Barnard}}, \ and\ \bibinfo {author} {\bibfnamefont {J.~A.~C.}\ \bibnamefont
  {Sandberg}},\ }\href@noop {} {\emph {\bibinfo {title} {{The think aloud
  method: A practical guide to modeling cognitive processes}}}}\ (\bibinfo
  {publisher} {Academic Press},\ \bibinfo {address} {London},\ \bibinfo {year}
  {1994})\BibitemShut {NoStop}%
\bibitem [{\citenamefont {Taylor}\ and\ \citenamefont
  {Dionne}(2000)}]{Taylor2000}%
  \BibitemOpen
  \bibfield  {author} {\bibinfo {author} {\bibfnamefont {K.~Lynn}\ \bibnamefont
  {Taylor}}\ and\ \bibinfo {author} {\bibfnamefont {Jean-Paul}\ \bibnamefont
  {Dionne}},\ }\bibfield  {title} {\enquote {\bibinfo {title} {Accessing
  problem-solving strategy knowledge: The complementary use of concurrent
  verbal protocols and retrospective debriefing},}\ }\href {\doibase
  http://dx.doi.org/10.1037/0022-0663.92.3.413} {\bibfield  {journal} {\bibinfo
   {journal} {Journal of Educational Psychology}\ }\textbf {\bibinfo {volume}
  {92}},\ \bibinfo {pages} {413--425} (\bibinfo {year} {2000})}\BibitemShut
  {NoStop}%
\bibitem [{\citenamefont {Pate}\ and\ \citenamefont {Miller}(2011)}]{Pate2011}%
  \BibitemOpen
  \bibfield  {author} {\bibinfo {author} {\bibfnamefont {Michael~L.}\
  \bibnamefont {Pate}}\ and\ \bibinfo {author} {\bibfnamefont {Greg}\
  \bibnamefont {Miller}},\ }\bibfield  {title} {\enquote {\bibinfo {title} {A
  descriptive interpretive analysis of studentsÕ oral verbalization during the
  use of thinkÐaloud pair problem solving while troubleshooting},}\ }\href
  {\doibase 10.5032/jae.2011.01107} {\bibfield  {journal} {\bibinfo  {journal}
  {Journal of Agricultural Education}\ }\textbf {\bibinfo {volume} {52}},\
  \bibinfo {pages} {107--119} (\bibinfo {year} {2011})}\BibitemShut {NoStop}%
\bibitem [{\citenamefont {Bronfenbrenner}(1977)}]{Bronfenbrenner1977}%
  \BibitemOpen
  \bibfield  {author} {\bibinfo {author} {\bibfnamefont {Urie}\ \bibnamefont
  {Bronfenbrenner}},\ }\bibfield  {title} {\enquote {\bibinfo {title} {Toward
  an experimental ecology of human development},}\ }\href {\doibase
  http://dx.doi.org/10.1037/0003-066X.32.7.513} {\bibfield  {journal} {\bibinfo
   {journal} {American Psychologist}\ }\textbf {\bibinfo {volume} {32}},\
  \bibinfo {pages} {513--531} (\bibinfo {year} {1977})}\BibitemShut {NoStop}%
\end{thebibliography}%

\end{document}